\documentclass[final,5p,times,twocolumn,authoryear,linenumbers]{elsarticle}
\usepackage{cuted}
\usepackage{graphicx}
\usepackage{multirow}
\usepackage{txfonts}
\usepackage{chngcntr}
\usepackage{natbib}
\usepackage{url}
\usepackage{hyperref}
\usepackage{longtable}
\usepackage{tabularx}
\usepackage{amssymb}
\usepackage{lipsum}
\usepackage{array}
\usepackage{rotating}
\usepackage{changepage}
\usepackage{lineno}
\usepackage{booktabs}
\usepackage{subcaption}

\journal{Journal of High Energy Astrophysics}

\begin{document}

\begin{frontmatter}

\title{Unveiling the Accretion Architecture of GS 1354--64: A Broadband X-ray and Polarimetric Study of its 2025--2026 Outburst}

\author[first,second]{Kaushik Chatterjee\fnref{star}}
\affiliation[first]{organization={South-Western Institute for Astronomy Research, Yunnan University},
            addressline={University Town, Chenggong},
            city={Kunming},
            postcode={650500},
            state={Yunnan},
            country={People's Republic of China}}	
\affiliation[second]{Organization={Key Laboratory of Survey Science of Yunnan Province, Yunnan University}, 
	    city={Kunming}, 
	    state={Yunnan}, 
	    postcode={650500}, 
	    country={People's Republic of China}}	    

\fntext[star]{Email Address: \href{mailto:mails.kc.physics@gmail.com}{mails.kc.physics@gmail.com} \& \href{mailto:kaushik@ynu.edu.cn}{kaushik@ynu.edu.cn}}

\begin{abstract}

A comprehensive multi-satellite analysis of the previously dynamically confirmed black hole candidate GS 1354--64 during its highly active 2025--2026 outburst is presented here. By leveraging simultaneous broadband spectroscopy, fast timing, and X-ray polarimetry, I map the evolution and geometry of the accretion flow. During the initial Compton-dominated hard state rise, strong, narrow Type-C quasi-periodic oscillations (QPOs) evolving from 0.08~Hz to 1.73~Hz with high fractional RMS ($\sim 8 - 10\%$) are detected. A strong anti-correlation between the QPO frequency and the inner disk radius ($R_{\rm in}$), alongside a linear correlation with the coronal photon index ($\Gamma$) are demonstrated. This supports the robust evidence that the inward migration of the disk drives the precessional frequency higher while efficiently cooling the corona. Concurrently, an anomalous, highly ionized (Fe XXVI) equatorial disk wind is observed, suggesting strong localized radiation pressure prior to the state transition. As the source entered a luminous, thermal-dominated soft state, individual epoch reflection modeling indicated a rapidly spinning black hole ($a_* \approx 0.94 - 0.99$) at an inclination of $i \approx 45^\circ - 53^\circ$. Subsequent joint spectral modeling across these reflection-dominated epochs refined these measurements, yielding a strictly anchored inclination of $i \approx 51^\circ$ and a near-maximal spin ($a_* \approx 0.998$) with the inner accretion disk rigidly pinned to the innermost stable circular orbit (ISCO). During this soft phase, the QPOs shifted to higher frequencies ($\sim 4.7 - 6.2$~Hz) and became relatively weak and broad (RMS dropping to $\sim 3 - 6\%$). Joint spectro-polarimetric modeling of this exact epoch reveals a steep, energy-increasing polarization degree, ruling out the possible existence of a spherical or lamppost geometry in favor of a highly asymmetric, radially extended slab-like corona. It is demonstrated that the suppression of the high-RMS macroscopic QPO, combined with the strict energy-independence of the polarization angle, provides compelling evidence that the corona stabilized into an extended, symmetric slab---structurally consistent with the collapse of the puffed-up spherical corona observed in the hard state.

\end{abstract}


\end{frontmatter}

\section{Introduction}
\label{sec:intro}

Compact objects, such as neutron stars and black holes (BHs), represent the end products of massive stellar evolution. Their intense gravitational pull prevents particles or radiation from escaping, meaning they can only be detected via the electromagnetic radiation emitted by matter accreting onto them. Most galactic black hole candidates (BHCs) reside in close binary systems where a companion star acts as a mass donor. Matter accreted either via stellar winds or Roche lobe overflow swirls towards the black hole, forming an accretion disk due to the conservation of angular momentum. The emission from these systems is primarily composed of thermal and non-thermal radiation. An optically thick and geometrically thin standard disk \citep{Shakura1973} is responsible for thermal blackbody radiation, generated by the conversion of gravitational potential energy into thermal energy as matter accreted inwards \citep{Frank1982}. Surrounding this is a hot Corona, a repository of high-energy electrons that intercepts soft photons from the accretion disk and up-scatters them to higher energies via inverse Comptonization \citep{Sunyaev1980, Chakrabarti1995}. This process produces the hard, non-thermal power-law tail observed in the spectra of accreting BHs.

Many BHCs in low-mass X-ray binaries are transient in nature, occasionally undergoing outbursts characterized by significant variability in both their temporal and spectral states \citep{Tomsick2000, Remillard2006}. During an active outburst phase, these objects typically transit through several distinct spectral states, tracing a characteristic ``q''-shaped loop in the Hardness-Intensity Diagram (HID). The canonical states include the hard state (HS), hard-intermediate state (HIMS), soft-intermediate state (SIMS), and soft state (SS) \citep{Remillard2006}. In the HS, the spectrum is dominated by the hard power-law component (photon index $\Gamma \approx 1.4-2.0$) with a high-energy cutoff near 100 keV, indicating a hot, optically thin corona. The system then transitions through the HIMS and SIMS at a roughly constant luminosity, eventually reaching the SS, where the thermal component from the inner accretion disk extends down to the innermost stable circular orbit (ISCO) and dominates the X-ray flux. Outbursts exhibit a strong hysteresis effect, wherein the hard-to-soft transition occurs at a higher luminosity than the subsequent soft-to-hard return transition. The outburst decay from the peak is largely controlled by X-ray irradiation of the outer disk, transitioning from a viscous decay stage to an irradiation-controlled stage governed by an inward-propagating cooling front \citep{Tetarenko2020}.

The timing properties of these systems provide an equally crucial diagnostic tool for mapping the inner accretion flow. The X-ray light curves of BHCs routinely display quasi-periodic oscillations (QPOs) superimposed on broad-band noise. Low-frequency QPOs (LFQPOs), with centroid frequencies ranging from mHz to $\sim 30$ Hz, are commonly observed during the hard and intermediate spectral states \citep{Remillard2006}. These LFQPOs are phenomenologically classified into Types A, B, and C \citep{Casella2005}. Type-C QPOs are strong, narrow features (quality factor $Q \ge 8$, up to 20\% rms amplitude) accompanied by flat-topped noise, and their centroid frequency monotonically increases with time as the source softens during the rising HS and HIMS \citep{Ingram2019, Chatterjee2024, Chatterjee2025a}. Type-B QPOs, peaking near 5--6 Hz with weak red noise, define the SIMS and are often associated with discrete jet ejections. Type-A QPOs are weak, broad features appearing just after the transition into the soft state.

In addition to X-ray spectral and timing signatures, the accretion configuration is intimately tied to the launching of jets and outflows. Matter ejected from the hot corona due to thermal or magnetic pressure forms jets \citep{Blandford1982, Fender2004}. Compact, steady radio jets are a hallmark of the hard state, whereas discrete, superluminal blobby ejections are observed as the system crosses the ``jet line'' in the intermediate states \citep{Fender1999, Dhawan2000}. Jet emission is entirely quenched in the soft state as the inner disk collapses inward and cools the jet base. Conversely, highly ionized, massive equatorial disk winds are preferentially observed during the soft state, potentially acting as an alternative mass-loss mechanism when the jet is quenched \citep{Neilsen2023}. However, recent high-resolution spectroscopy has also revealed ultra-fast outflows (UFOs) during the hard state, requiring intense magnetic driving mechanisms from the inner disk \citep{Miller2015}.

Recently, X-ray polarimetry has emerged as a groundbreaking diagnostic tool, offering independent geometric constraints on the inner accretion flow inaccessible through flux spectroscopy alone. In black hole systems, linear polarization primarily arises from standard electron (Thomson) scattering within the optically thick disk and the Comptonizing corona \citep{Schnittman2009}. The polarization degree (PD) and angle (PA) are highly sensitive to the spatial asymmetry of the scattering medium and viewing inclination \citep{Haardt1993, Poutanen1993}. While spherically symmetric clouds yield zero net polarization, slab-like or wedge-like geometries produce a distinct, energy-dependent increase in PD. The PA is dictated by the symmetry axis of the flow; a strictly constant, energy-independent PA implies a single global axis of symmetry, whereas macroscopic precessional wobbling naturally introduces complex energy-dependence in the polarization vector \citep{Ingram2015, Connors2020}. Observations by the Imaging X-ray Polarimetry Explorer (\textit{IXPE}) \citep{Weisskopf2022} have successfully broken such geometric degeneracies. For instance, \textit{IXPE} revealed polarization signatures in Cygnus X-1 during the hard state consistent with a corona extended in the plane of the disk \citep{Krawczynski2022}. A comprehensive multi-mission study of 12 black hole X-ray binaries further established a strong positive correlation between PD and the Comptonized photon fraction, highlighting significant polarization in sources like Cyg X-3 and IGR J17091-3624 \citep{Majumder2026}. Conversely, while an initial study of the transient Swift J151857.0--572147 suggested a low polarization degree in its soft state \citep{Mondal2024}, a rigorous re-analysis demonstrated that the polarization is actually statistically unconstrained (with an upper limit of $0.8\%$), completely consistent with the unpolarized emission expected from a standard thin accretion disk \citep{Ling2024}. 

Alongside polarimetry, advancements in high-resolution X-ray spectroscopy via microcalorimeters (e.g., \textit{XRISM}) allow for precise mapping of relativistically broadened reflection features to determine the inner disk radius and black hole spin ($a_*$) \citep{Garcia2014, Chatterjee2025b, Liu2026}. To interpret these complex phenomena, phenomenological models like the Relativistic Precession Model (RPM) \citep{Stella1998, Ingram2009} attribute LFQPOs to Lense-Thirring precession of the misaligned hot inner flow. Conversely, physical models like the Two-Component Advective Flow (TCAF) \citep{Chakrabarti1995} extract mass accretion rates by treating the Compton cloud as a centrifugal boundary layer (CENBOL) \citep{Debnath2014, Mondal2014, Molla2016, Chatterjee2020, Chatterjee2021, Chatterjee2023, Chatterjee2025b}.

The transient low-mass X-ray binary GS 1354-64 is a dynamically confirmed black hole system ($5.7 \pm 0.3 M_{\odot}$, $d = 0.7_{-0.3}^{+3.4}$ kpc; \citealt{Casares2009, Gandhi2019}). Its previous outbursts in 1997 and 2015 were "failed" transitions confined to hard states, where deep \textit{NuSTAR} reflection modeling indicated a highly truncated disk and a near-maximal spin ($a_* > 0.98$) \citep{Koljonen2016, ElBatal2016}. However, its bright 2025--2026 outburst marked a definitive departure, completing a full hard-to-soft state transition \citep{Negoro2025, Negoro2026a, Negoro2026b}. Early high-resolution \textit{XRISM} and \textit{NuSTAR} observations during the rising hard state revealed unblended broad iron lines and signatures of ultra-fast outflows \citep{Adegoke2026, Liu2026}. Crucially, \textit{IXPE} observed the system in an intermediate state following a stalled transition. Analyzing this specific epoch, \citet{Ravi2026} discovered a $4.0\% \pm 0.2\%$ PD that increased with energy, alongside a $\sim 5$ Hz Type-C QPO, concluding the presence of a radially extended corona at that time.

While \citet{Ravi2026} provided a detailed polarimetric view of a single intermediate-state snapshot, the overarching dynamic evolution of the accretion flow throughout the entire outburst remained unexplored. In this paper, I present a comprehensive spectral, temporal, and polarimetric analysis spanning the full 2025--2026 outburst of GS 1354-64, capturing its complete hard-to-soft-to-hard state transition. By uniting broadband \textit{NuSTAR} constraints across ten distinct epochs with the \textit{IXPE} polarimetry, and utilizing dead-time free cross-spectra to trace the unsuppressed evolution of the LFQPOs, I track the macroscopic geometric evolution of the corona and inner disk. This broad baseline explicitly tests macroscopic precession models against radial geometric oscillations to determine the true structural nature of the soft-state corona. The paper is organized as follows: In Section 2, I describe the observations and data reduction processes strictly for the \textit{NuSTAR} and \textit{IXPE} observatories. In Section 3, I detail the spectral and temporal data analysis techniques. In Section 4, I report the results of these investigations. In Section 5, I discuss the physical implications of these findings in the context of accretion flow dynamics, LFQPO evolution, and coronal geometry. Finally, in Section 6, I summarize the results and draw concluding remarks.

\section{Observations and Data Reduction}
\label{sec:data}

The 2025--2026 outburst of GS 1354-64 was extensively monitored by a multitude of X-ray observatories. To investigate the broadband spectral and temporal evolution of the source across its state transitions, I utilize data from \textit{NuSTAR}, and \textit{IXPE}. The comprehensive observation logs for all instruments used in this study, including the precise Start UT, End UT, Start MJD, End MJD, and Average MJD, are summarized in Table \ref{tab:obs_log}.

\begin{table*}[ht!]
\centering
\scriptsize
\caption{Comprehensive Summary of Space Observatory Log covering the 2025--2026 Outburst of GS 1354-64.}
\label{tab:obs_log}
\renewcommand{\arraystretch}{1.2}
\setlength{\tabcolsep}{6pt}
\begin{tabular}{llccccc}
\hline\hline
Instrument & ObsID / Exposure ID & Start UT & End UT & Start MJD & End MJD & Avg MJD \\
\hline
\multicolumn{7}{l}{\textbf{NuSTAR}} \\
\hline
NuSTAR & 91201302002 & 2026-01-13 08:05:55 & 2026-01-13 20:50:39 & 61053.33745 & 61053.86851 & 61053.60298 \\
NuSTAR & 91201302004 & 2026-01-21 07:05:51 & 2026-01-21 19:00:39 & 61061.29573 & 61061.79212 & 61061.54393 \\
NuSTAR & 91201302006 & 2026-01-24 11:26:56 & 2026-01-24 21:45:39 & 61064.47704 & 61064.90670 & 61064.69187 \\
NuSTAR & 91202308002 & 2026-01-30 02:40:13 & 2026-01-30 13:11:17 & 61070.11127 & 61070.54951 & 61070.33039 \\
NuSTAR & 91202311002 & 2026-02-11 18:17:23 & 2026-02-11 22:28:03 & 61082.76207 & 61082.93616 & 61082.84911 \\
NuSTAR & 91202311004 & 2026-02-14 01:57:03 & 2026-02-14 06:07:17 & 61085.08129 & 61085.25506 & 61085.16818 \\
NuSTAR & 91202311006 & 2026-02-20 01:03:17 & 2026-02-20 05:12:21 & 61091.04395 & 61091.21691 & 61091.13043 \\
NuSTAR & 91202311008 & 2026-02-22 13:28:41 & 2026-02-22 17:38:38 & 61093.56159 & 61093.73517 & 61093.64838 \\
NuSTAR & 91202311010 & 2026-02-27 07:57:39 & 2026-02-27 12:07:05 & 61098.33171 & 61098.50492 & 61098.41832 \\
NuSTAR & 91202311012 & 2026-03-18 09:51:54 & 2026-03-18 14:00:52 & 61117.41105 & 61117.58394 & 61117.49749 \\
\hline
\multicolumn{7}{l}{\textbf{IXPE}} \\
\hline
IXPE   & 05002101   & 2026-02-07 06:53:56 & 2026-02-12 20:28:19 & 61078.28745 & 61083.85299 & 61081.07022 \\
\hline\hline
\end{tabular}
\end{table*}

\subsection{Data Reduction}
All data reduction and analysis procedures were performed using the standard \texttt{HEASoft} software suite (version 6.35.1).

\subsubsection{NuSTAR}
\label{subsec:nustar}
Cleaned event files were generated using the \texttt{nupipeline} task within the \texttt{NuSTARDAS} software, applying the relevant calibration database (CALDB; \texttt{version:20260421}). I extracted the source spectra and light curves from a circular region with a radius of 80 arcsec centered on the source coordinates. Background spectra were extracted from source-free circular regions with also the same radius of 80 arcsec, on the same detectors. Science-ready spectra and light curves were subsequently generated using the \texttt{nuproducts} command. To ensure adequate statistics for $\chi^2$ minimization, the spectra were grouped to a minimum of 30 counts per bin using the \texttt{grppha} task.

\subsubsection{IXPE}
\label{subsec:ixpe}
For the polarimetric analysis of GS 1354-64, pre-cleaned Level 2 event files for the Detector Units were acquired from the public \href{https://heasarc.gsfc.nasa.gov/cgi-bin/W3Browse/w3browse.pl}{HEASARC archive}. Data reduction, calibration, and filtering were executed utilizing the \texttt{ixpeobssim} software suite \citep[v31.0.3;][]{Baldini2022} in conjunction with standard \texttt{FTOOLS} routines, referencing the instrumental calibration database (CALDB \texttt{version:20260331}). The overall processing and handling of the \textit{IXPE} data followed the standard prescriptions detailed in \citet{DiMarco2026}.

Source-specific events were isolated using the \texttt{xpselect} tool by defining a standard extraction circle with a radius of $60^{\prime\prime}$ anchored at the object's centroid coordinates. This extraction size is well within the recommended spatial selection bounds for \textit{IXPE} point sources \citep{DiMarco2026}. Following the strict recommendations for bright sources outlined by \citet{DiMarco2023}, background subtraction was not performed; for bright sources ($> 2 \rm\ cps/arcmin^2$), background subtraction from the same field should be avoided because the background region is inherently dominated by the source itself.

To extract the polarimetric quantities, two parallel methodologies were employed. First, for the model-independent analysis, I implemented the unweighted \texttt{PCUBE} algorithm natively available within \texttt{ixpeobssim} \citep{Kislat2015}. This extracts the polarization degree ($PD$) and polarization angle ($PA$) from linear combinations of the Stokes profiles, automatically correcting for the energy-dependent modulation factor ($\mu$). Second, for the model-dependent joint spectro-polarimetric analysis in \texttt{XSPEC}, science-ready Stokes $I$, $Q$, and $U$ spectra were generated. To optimize polarization sensitivity and correct for instrumental deviations at higher energies, these spectra were extracted utilizing the weighted approach \citep{DiMarco2022}, and the ancillary response functions (ARFs) were explicitly corrected for vignetting using the \texttt{IXPECALCARF} tool.

\section{Data Analysis}
\label{sec:analysis}

A detailed spectral and timing analysis of GS 1354-64 is performed with broadband and polarimetric data. I discuss the detailed analysis methodology in the subsequent subsections.

\subsection{Spectral Analysis}
\label{subsec:spectral_analysis}

\begin{figure}[!h]
\centering
\textbf{Unfolded Model Fitted Spectra}\par\medskip
\vbox{
\includegraphics[width=8.0truecm,angle=0]{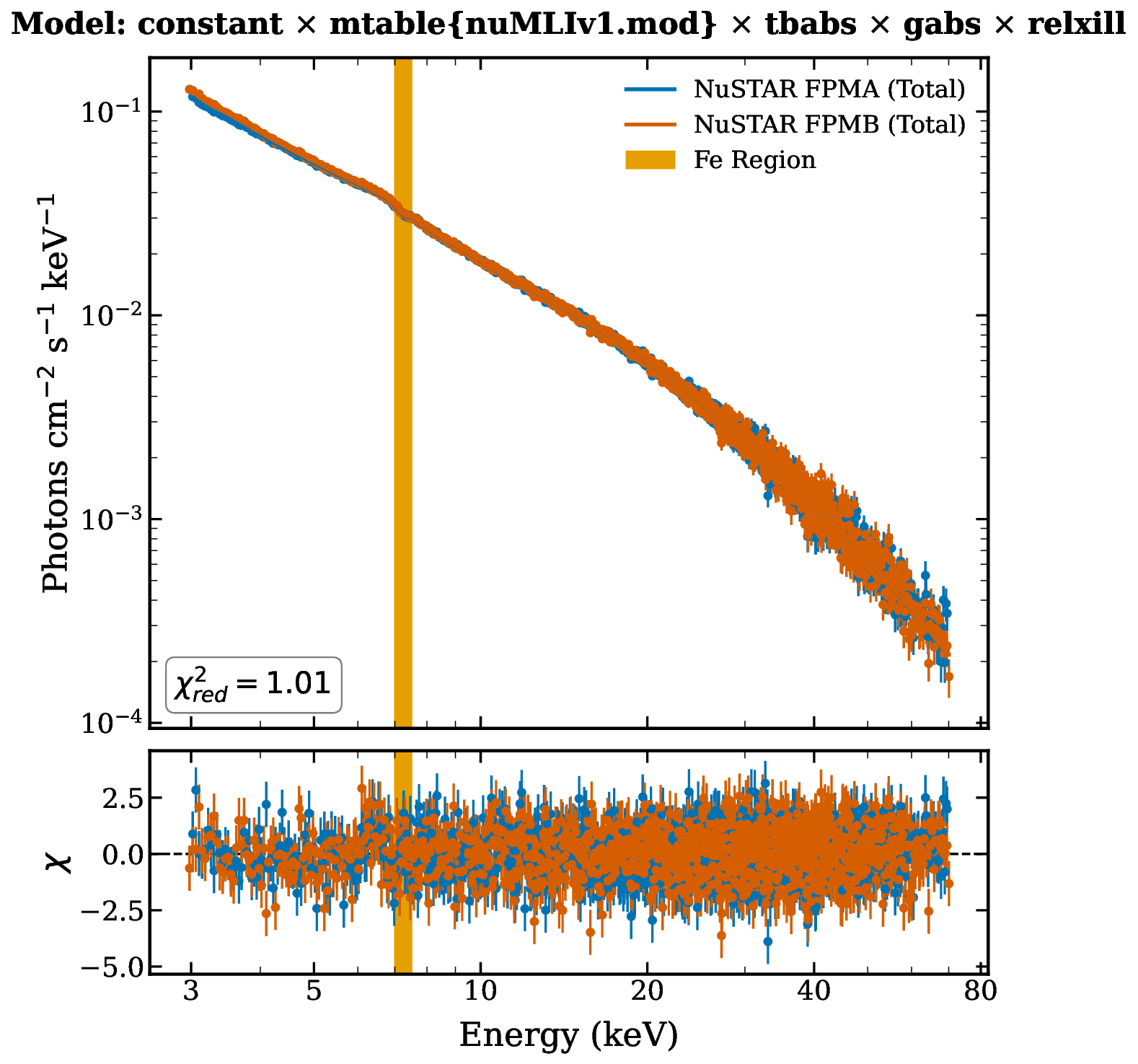}}
\includegraphics[width=8.0truecm,angle=0]{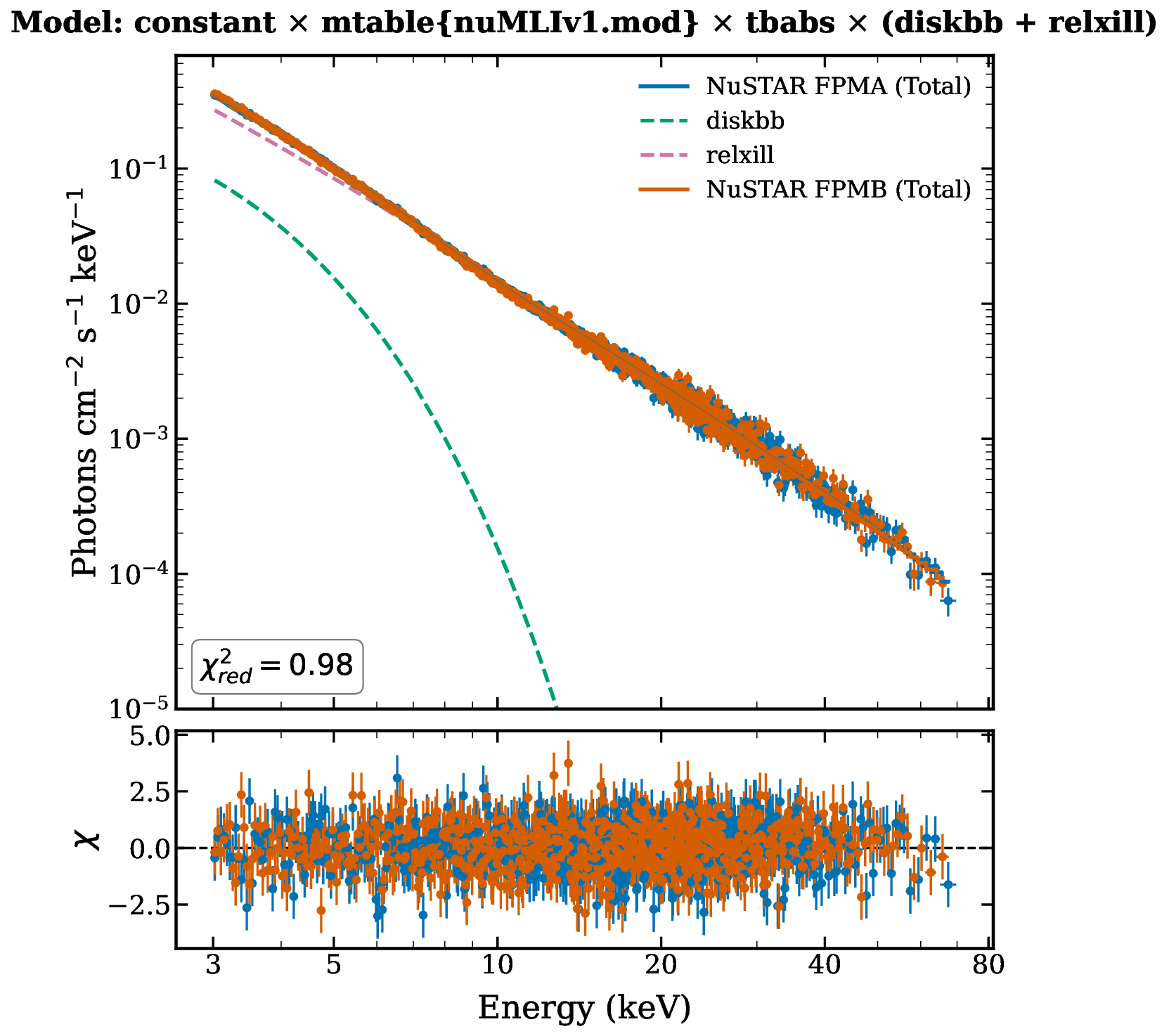}
\includegraphics[width=8.0truecm,angle=0]{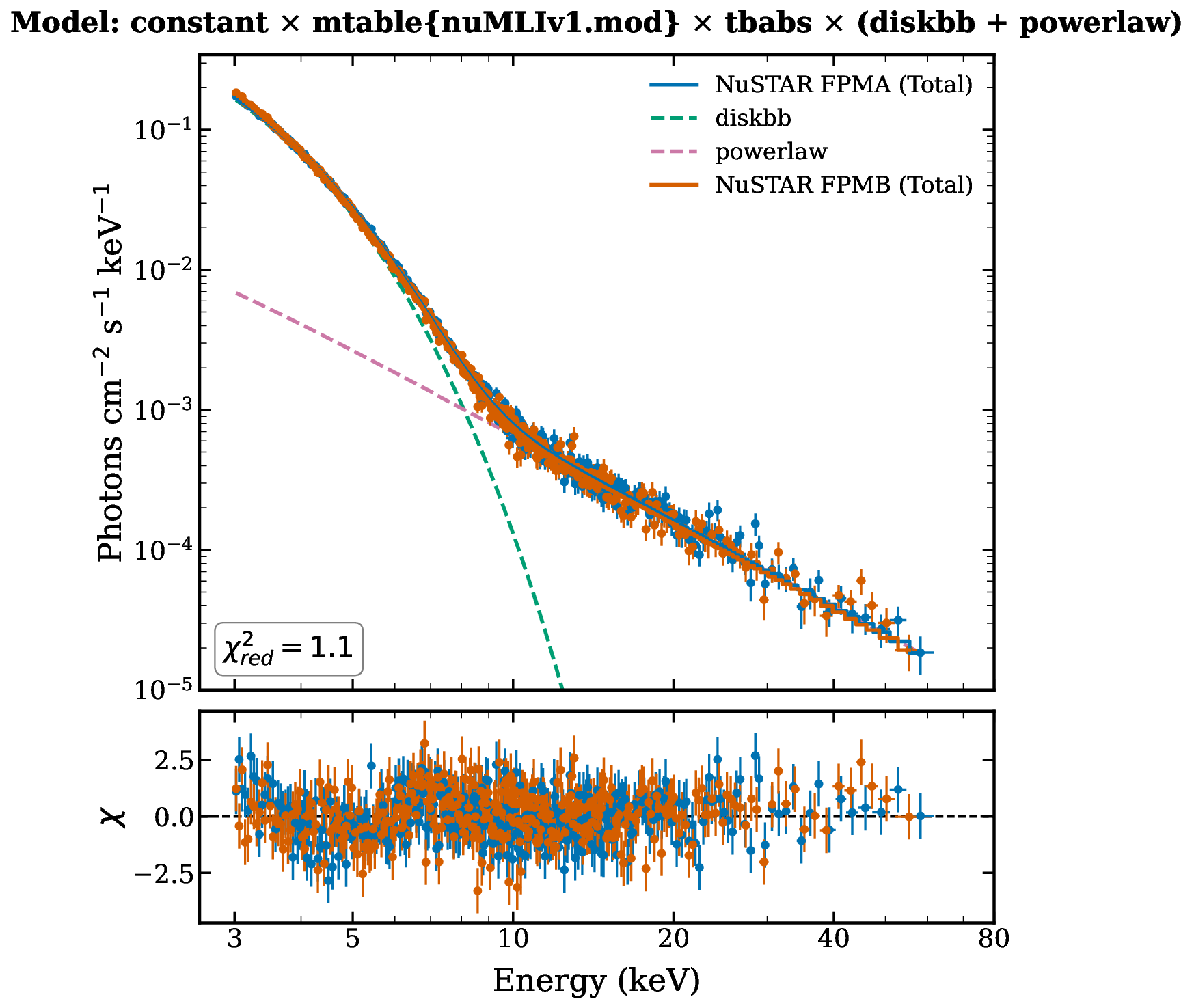}
\caption{Best-fitted unfolded broadband spectra in $3-70 ~keV$ energy band using \textit{NuSTAR} data for different sets of model combinations.}
\label{fig:unfolded_spec}
\end{figure}

The broadband spectral analysis of the \textit{NuSTAR} observations was performed in the 3--70 keV energy range using the \texttt{XSPEC} spectral fitting package \citep[version 12.15.0;][]{Arnaud1996}. The data above 70 keV had a lot of noise and was decided to leave out. During the initial spectral extraction, a large discrepancy between the FPMA and FPMB detectors was noted below 7 keV. Following direct consultation with the \textit{NuSTAR} Science Operations Center (SOC), this soft-excess discrepancy was confirmed to be a known instrumental artifact caused by a tear in the thermal blanketing (MLI) affecting the FPMB module. To rigorously correct this, the \texttt{nuMLIv1.mod} table model was applied to the continuum. The FPMA MLI fraction was strictly frozen to 1.0, while the FPMB MLI fraction was allowed to vary. Across all epochs, the FPMB MLI fraction consistently evaluated to between 0.70 and 0.95, which, alongside the complete resolution of the 3--4 keV residuals, successfully verified the MLI tear as the sole cause of the discrepancy. Given that the spectra were grouped to a minimum of 30 counts per bin during data reduction, I utilized standard $\chi^2$ minimization to evaluate the goodness-of-fit for all models.

To account for the interstellar absorption along the line of sight towards GS 1354-64, I employed the \texttt{tbabs} model in all of the fits. Because the unmodeled instrumental soft excess previously induced artificial degeneracies in the absorption column, implementing the MLI correction allowed the line-of-sight column density ($N_H$) to be strictly frozen to the established Galactic predicted value of $0.70 \times 10^{22}\ \rm cm^{-2}$ across all observations. The interstellar medium (ISM) abundances were set according to \citet{Wilms2000}, and the photoelectric cross-sections were adopted from \citet{Verner1996}. Furthermore, a multiplicative \texttt{constant} component was included in every model iteration to account for the slight cross-calibration uncertainties between the two \textit{NuSTAR} focal plane modules. This constant was fixed to unity for the FPMA detector and allowed to vary freely for the FPMB detector.

To properly characterize the broadband X-ray continuum, the broad Iron K$\alpha$ line complex, and the high-energy Compton hump, I adopted a systematic modeling framework based on the relativistic reflection suite \texttt{relxill}. Depending on the specific spectral state and the physical features present during a given observation, I applied one of the following three model combinations:

\begin{enumerate}

    \item \texttt{constant * mtable\{nuMLIv1.mod\} * tbabs * gabs * relxill}: This model combination was utilized during the initial rising hard and hard-intermediate states. In these regimes, the spectrum is dominated by the primary Comptonized emission from the corona and its reflection off the inner accretion disk. Furthermore, significant residuals indicated the presence of narrow absorption features superimposed on the reflection continuum---indicative of highly ionized disk winds (e.g., Fe XXVI)---necessitating the inclusion of the multiplicative Gaussian absorption line model, \texttt{gabs} \citep{Arnaud1996}.
    
    \item \texttt{constant * mtable\{nuMLIv1.mod\} * tbabs * (diskbb + relxill)}: As the outburst transitioned toward the soft state, the accretion disk moved inward and heated up. For these observations, I applied the multi-color disk blackbody component \citep[\texttt{diskbb};][]{Mitsuda1984, Makishima1986} to account for the direct thermal emission from the inner accretion flow emerging at the lower end of the \textit{NuSTAR} energy range.
    
    \item \texttt{constant * mtable\{nuMLIv1.mod\} * tbabs * (diskbb + powerlaw)}: For the final observation characterizing the faint decay phase, the overall X-ray flux dropped substantially. Consequently, the relativistic reflection features faded into the background noise. Applying the \texttt{relxill} model to this low-luminosity epoch resulted in extreme statistical degeneracies as the model chased noise. Therefore, the reflection component was removed entirely, and the continuum was successfully modeled using a simple \texttt{powerlaw} to capture the hardening of the primary emission.

\end{enumerate}

One of the best-fitted spectra from all three model combinations are shown in Figure \ref{fig:unfolded_spec}.

For the \texttt{relxill} component across all applicable models, I carefully monitored parameters such as the inner emissivity index ($q_{\rm in}$), the dimensionless black hole spin ($a$), the inclination angle ($i$), the ionization parameter ($\log \xi$), and the reflection fraction ($R_f$) to track the geometric and dynamic evolution of the inner accretion flow throughout the outburst.

While other flavors of the reflection suite, such as \texttt{relxillcp}, provide a more physically self-consistent treatment of the Comptonized continuum by parameterizing the electron temperature ($kT_e$), the base \texttt{relxill} model was deliberately chosen to prevent severe parameter degeneracy. Because the NuSTAR data above 70 keV was dominated by background noise and subsequently ignored, the high-energy rollover of the coronal spectrum cannot be observationally constrained. Consequently, the cutoff energy in our models was fixed well outside the observable bandpass at $E_{cut} = 300$ keV. Employing a model with a free electron temperature parameter is unsupported by the available energy range. The base \texttt{relxill} model sufficiently maps the relativistic broadening of the Fe K$\alpha$ complex and the reflection continuum necessary to constrain the inner accretion geometry without over-parameterizing the spectral fits.

\begin{figure*}[ht!]
    \centering
    \includegraphics[width=0.85\textwidth]{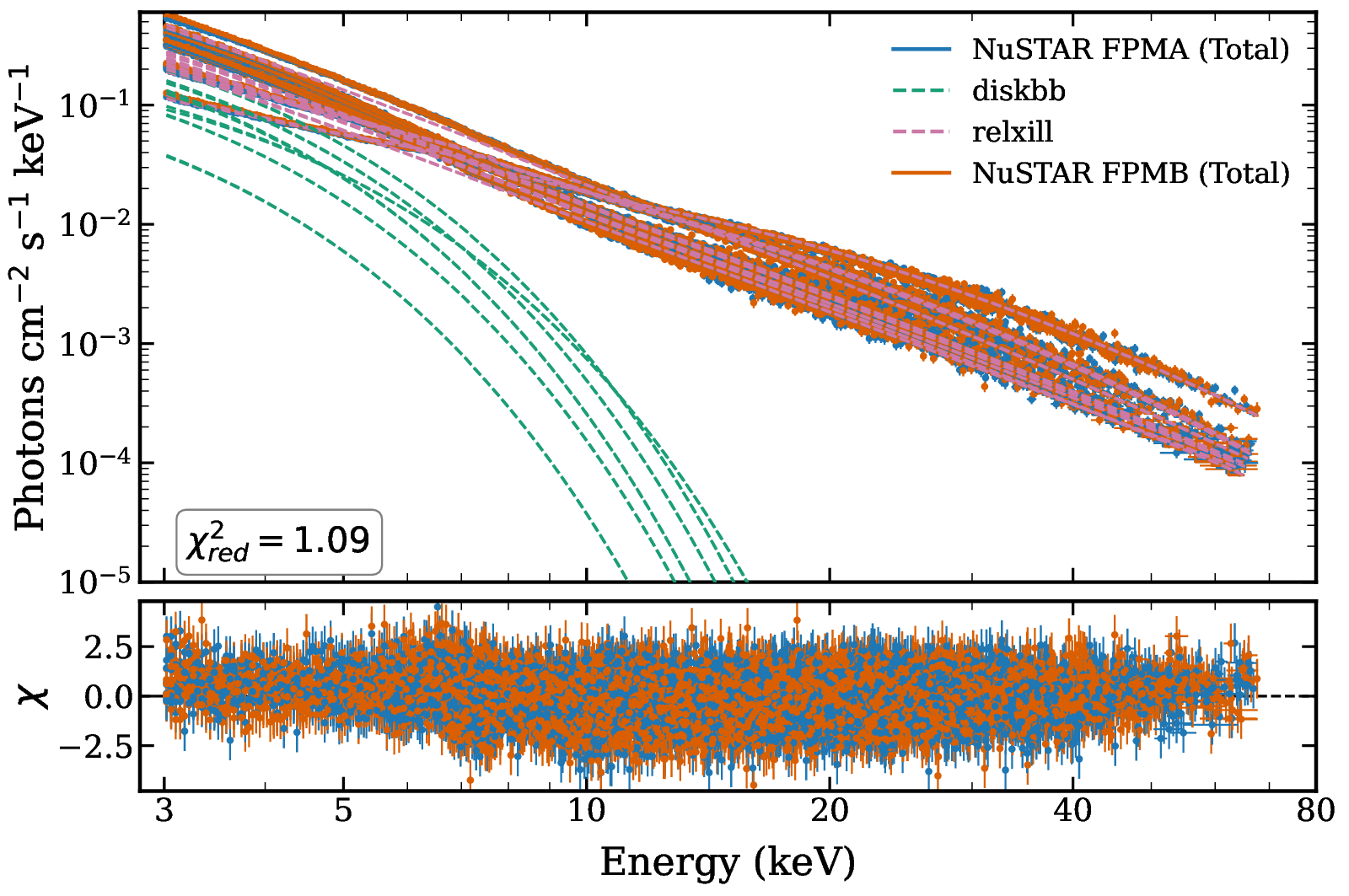}
    \caption{Best-fit unfolded broadband spectra and $\chi$ residuals for the joint spectral analysis across the active reflection-dominated states of GS 1354--64. The \textit{NuSTAR} FPMA (blue) and FPMB (orange) data are shown alongside the tied additive components: the thermal accretion disk (\texttt{diskbb}; dashed green) and the relativistically broadened coronal reflection (\texttt{relxill}; dashed pink). By physically tying the invariant geometric parameters (spin, inclination, and iron abundance) across all epochs, the joint fit achieves a statistically robust global minimum ($\chi^2_\nu = 1.09$) while eliminating individual observation degeneracies.}
    \label{fig:joint_fit_spec}
\end{figure*}

To rigorously constrain the system's static architecture and break potential geometric degeneracies, a joint spectral fit was subsequently executed across all active reflection-dominated epochs. During this joint modeling, the intrinsic physical parameters that are not expected to vary on outburst timescales---specifically the black hole spin ($a_*$), inner disk inclination ($i$), and iron abundance ($A_{\rm Fe}$)---were formally tied across all datasets. Conversely, state-dependent parameters such as the thermal disk temperature ($T_{\rm in}$), coronal photon index ($\Gamma$), ionization parameter ($\log \xi$), and reflection fraction ($R_f$) were allowed to vary freely for each individual observation. The best-fitted joint observation unfolded spectra is shown in Figure \ref{fig:joint_fit_spec}.

\subsection{Timing Analysis}
\label{subsec:temporal_analysis}

\begin{figure}[ht!]
    \centering
    \includegraphics[width=\columnwidth]{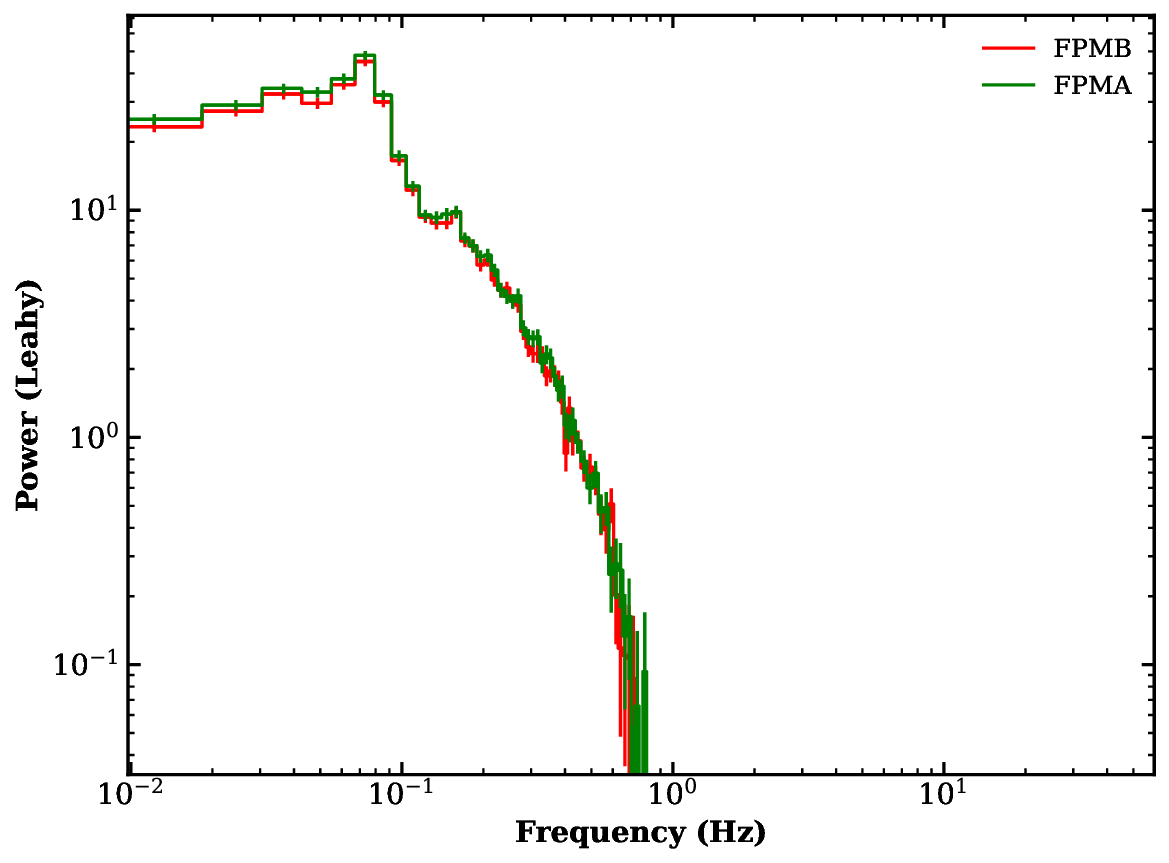}
    \caption{Leahy-normalized power density spectra (PDS) for the \textit{NuSTAR} FPMA (green) and FPMB (red) detectors. The expected Poisson white noise level has been subtracted. Both spectra exhibit the characteristic broad-band noise components associated with the accretion flow.}
    \label{fig:pds_fpma_fpmb}
\end{figure}

\begin{figure}[ht!]
    \centering
    \includegraphics[width=\columnwidth]{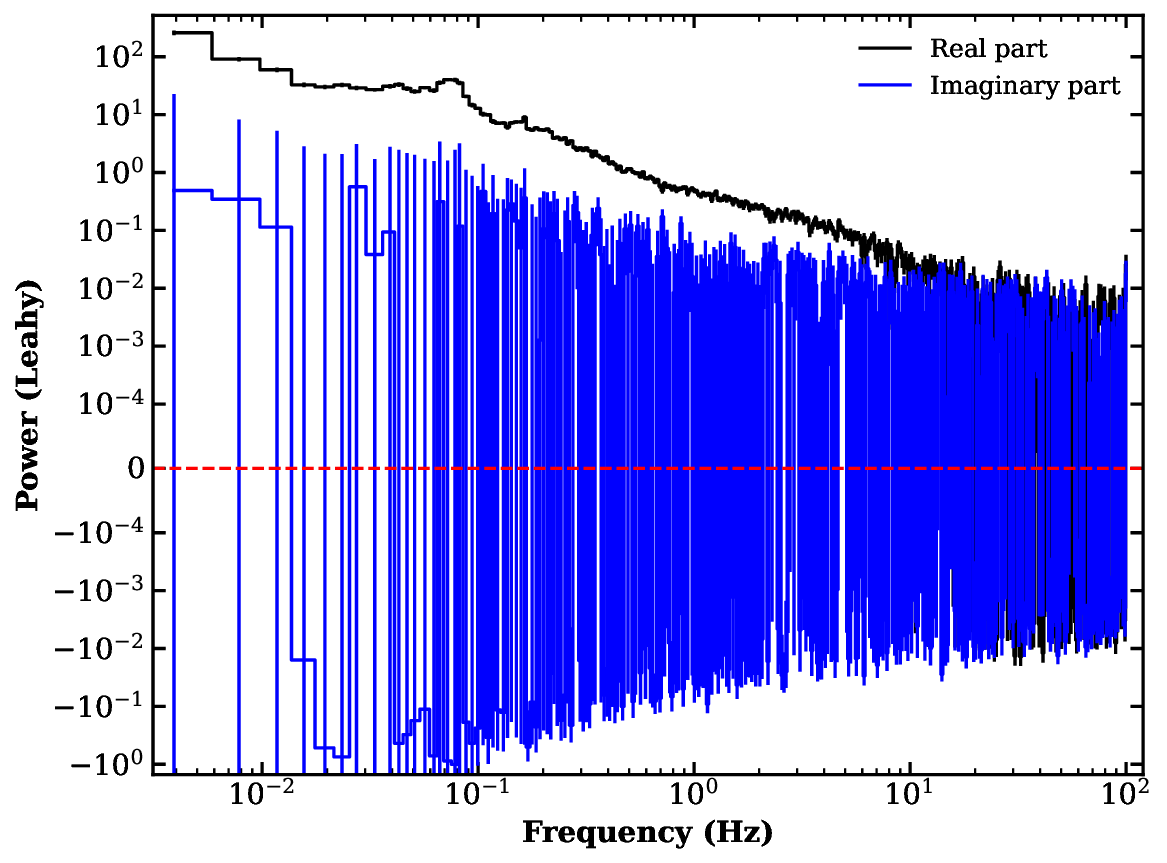}
    \caption{\textit{NuSTAR} cross-spectrum computed between the FPMA and FPMB event streams to eliminate instrumental dead-time effects. The real part (cospectrum) is shown in black, and the imaginary part (sinespectrum) is overlaid in blue. The sinespectrum fluctuates consistently around zero (red dashed line), confirming the absence of systematic instrumental phase lags between the two modules.}
    \label{fig:cross_spectrum}
\end{figure}

To investigate the fast temporal variability of GS 1354-64 during the outburst, I extracted high-resolution ($\Delta t = 0.005$ s) light curves from the \textit{NuSTAR} event files. However, timing analysis using individual \textit{NuSTAR} focal plane modules is profoundly limited by the instrument's large, non-paralyzable dead time ($\approx 2.5$ ms). Dead time suppresses the variance of the observed photon counting statistics, causing the high-frequency Poisson noise floor to drop significantly below the theoretical Leahy level of 2.0 \citep{Bachetti2015}.

This instrumental distortion is starkly evident in Figure \ref{fig:pds_fpma_fpmb}, which presents the individual, white-noise-subtracted power density spectra (PDS) for the FPMA and FPMB detectors. While the low-frequency broad-band noise is visible, the power appears to abruptly ``cut off'' and vanish from the logarithmic plot just above $\sim 1$ Hz. This is not an intrinsic source feature. Rather, because the dead-time suppresses the actual high-frequency noise floor, subtracting the theoretical standard noise level results in mathematically negative power at higher frequencies, causing the data to drop off the log-log axes entirely.

To overcome this severe limitation and recover the full-range source variability, I employed the cross-spectrum technique. Because the dead-time processes in the FPMA and FPMB detectors are strictly independent, computing the cross-spectrum between their simultaneous, non-overlapping 16 s light curve segments averages the uncorrelated Poisson noise to precisely zero \citep{Bachetti2015, Mendez2024}. The results of this cross-correlation methodology are presented in Figure \ref{fig:cross_spectrum}, utilizing logarithmic rebinning ($f = 0.02$). Recently, I used this approach to overcome the same dead-time issue for NuSTAR data for the BHC MAXI J1803-298 in \citet{Chatterjee2026}.

The real part of the averaged cross-spectrum (the cospectrum) cleanly isolates the intrinsic variability of the source. Unlike the individual PDS, the cospectrum is entirely free from dead-time suppression, revealing that the true broad-band noise smoothly extends across the full frequency range up to the Nyquist limit, dropping off naturally rather than hitting an artificial instrumental wall. Furthermore, the imaginary part (sinespectrum) is shown tightly fluctuating around zero across the entire frequency range. Because both focal plane modules observed the identical broad energy band simultaneously, no astrophysical time lag is present. The strictly zero-centered sinespectrum robustly confirms the absence of systematic instrumental phase lags between the two detectors. This successfully validated, dead-time-free cospectrum now provides a pristine dataset, fully prepared for future multi-Lorentzian modeling of the source's intrinsic broad-band noise and quasi-periodic oscillations.

\begin{figure}[ht!]
    \centering
    \includegraphics[width=\columnwidth,height=8cm]{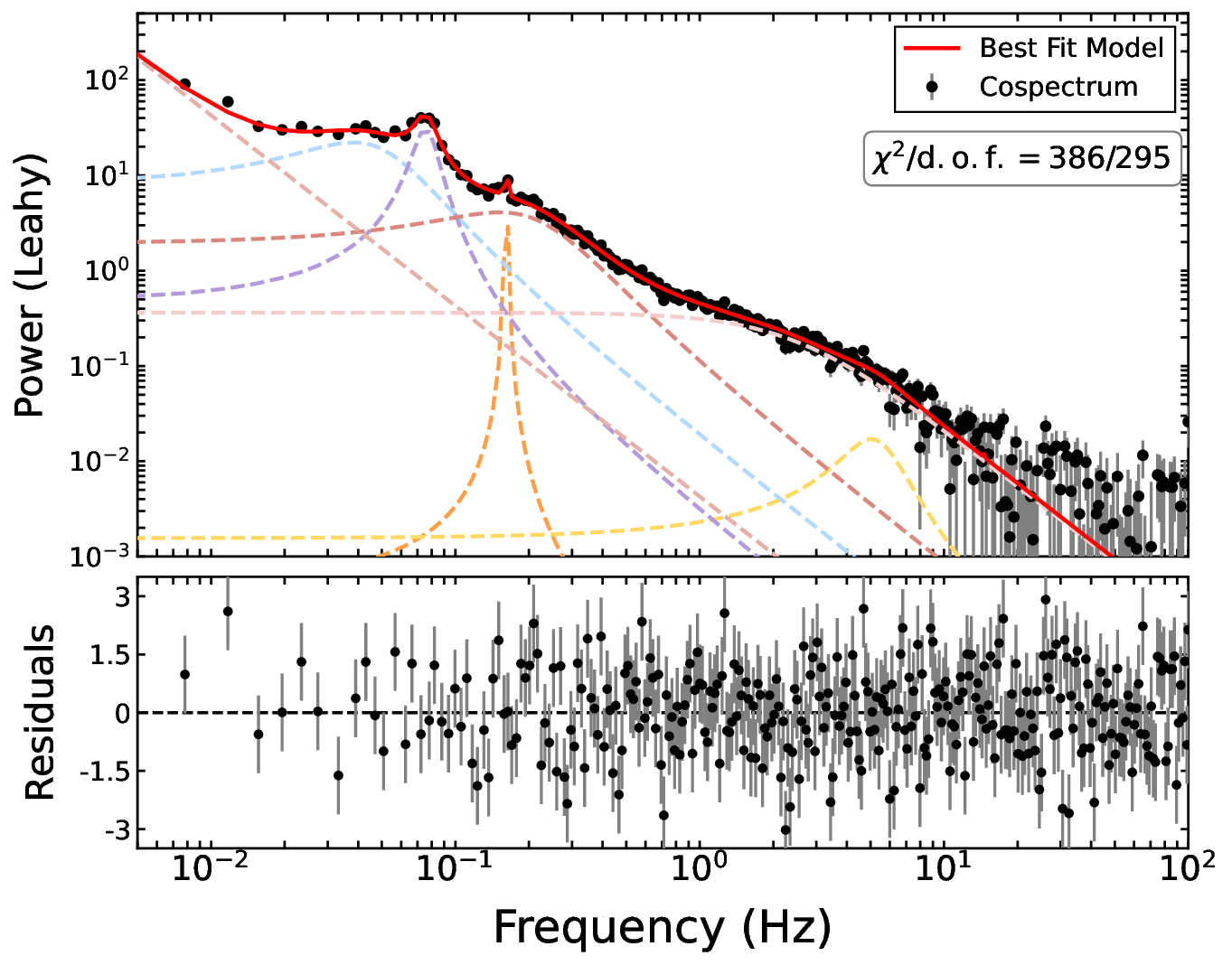}
    \caption{Best model fitted unfolded cospectrum using multiple Lorentzian models for the obs. ID. 91201302002. The contribution from each component could be noticed clearly.}
    \label{fig:pds_fit}
\end{figure}

Using the cospectrum, I have performed the model fitting in \texttt{XSPEC}. To achieve the best-fit, I used multiple Lorentzian profiles to take care for the contribution from the fundamental QPO, harmonics (if present), and the broadband noise. One of the best-fitted unfolded PDS (for observation ID 91201302002) is shown in the Figure \ref{fig:pds_fit}. 

During the highly dynamic hard-intermediate state (e.g., MJD 61061.54; ObsID 91201302004), the fundamental QPO exhibited a complex, broadened structure that could not be adequately captured by a single Lorentzian profile. Fitting this feature with a single component resulted in significant, structured wave-like residuals around the centroid frequency, indicative of either rapid frequency drift during the observation or a complex, dual-precessional geometry. To achieve a statistically acceptable fit and fully capture the signal variance, I introduced a second narrow Lorentzian component adjacent to the fundamental peak. This dual-peak modeling successfully flattened the residuals, confirming the physical presence of a split QPO track. A direct visual comparison of the single-peak versus dual-peak fitting residuals is provided in Appendix \ref{sec:appendix_qpo_split} in Figure \ref{fig:qpo_split_comparison}.

In addition to extracting time-averaged cosspectra, I utilized these continuous, dead-time-free data segments to compute the dynamical cospectrum. By isolating coherent astronomical variability in sequential 1024-second windows, this technique preserves critical time resolution. This allows us to visually track the structural evolution of QPOs throughout individual observations prior to executing the formal multi-Lorentzian time-averaged fitting.

\subsection{Polarization Analysis}
\label{subsec:polarization_analysis}

\begin{figure*}[h!]
    \centering
    \includegraphics[width=\textwidth,height=7cm]{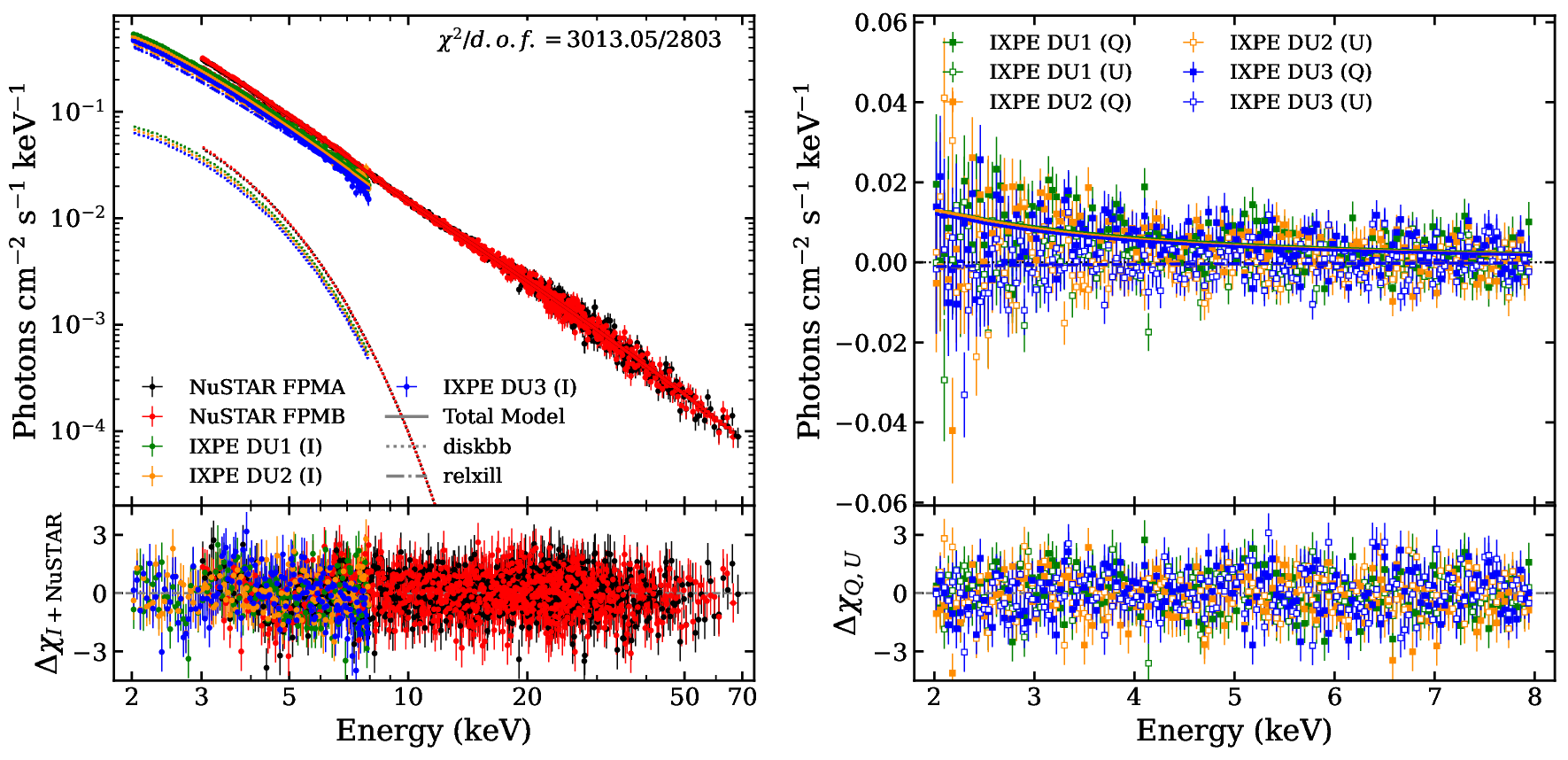}
	\caption{Joint multi-instrument spectro-polarimetric analysis of GS 1354-64 in a 2-column, 4-panel configuration. \textit{Left Column:} Broadband total intensity spectrum (Stokes $I$) plotted in standard energy flux units ($\rm Photons\ cm^{-2}\ s^{-2}\ keV^{-1}$) across the 2--70 keV range, merging \textit{NuSTAR} FPMA (black), FPMB (red), and \textit{IXPE} DU1 (green), DU2 (orange), and DU3 (blue) data. The best-fit total model is shown as a solid line, with individual component paths highlighted for \texttt{diskbb} (dotted line) and \texttt{relxill} (dash-dotted line). The lower panel provides the corresponding joint $\Delta\chi_{I+\rm NuSTAR}$ residuals. \textit{Right Column:} Energy-resolved linear polarization profile (Stokes $Q$ and $U$) over the high-resolution 2--8 keV linear X-axis, detailing filled squares for Stokes $Q$, open squares for Stokes $U$, their best-fit step lines, and the independent polarimetric $\Delta\chi_{Q,U}$ residuals.}
    \label{fig:joint_2col_spec}
\end{figure*}

To rigorously track the geometric orientation of the scattering media during the outburst, I conducted a comprehensive dual-methodology polarization analysis incorporating both model-independent algorithms and joint \texttt{XSPEC} spectro-polarimetric modeling.

First, to extract the model-independent polarimetric quantities for energy-resolved tracking, I implemented the \texttt{PCUBE} algorithm natively available within the \texttt{ixpeobssim} package. This algorithm utilizes a binned Stokes parameter method \citep{Kislat2015} to construct the $I$, $Q$, and $U$ photon counting cubes directly from the photoelectron track modulation patterns. The polarization degree ($PD$) and polarization angle ($PA$) are mathematically extracted from the linear combinations of these Stokes profiles, automatically subtracting the instrumental background and correcting for the energy-dependent modulation factor ($\mu$).

Second, to unify these polarimetric vectors with the physical continuum properties, I executed a joint broadband spectro-polarimetric fit targeting the luminous soft state window where \textit{IXPE} and \textit{NuSTAR} observations overlap. This was achieved by loading the \textit{NuSTAR} FPMA/FPMB data alongside the \textit{IXPE} Stokes $I$, $Q$, and $U$ event spectra from Detector Units 1, 2, and 3. To model the multi-dataset configuration, I extended the empirical framework by implementing a multiplicative polarization tracking network natively inside \texttt{XSPEC}: \texttt{constant * mtable\{nuMLIv1.mod\} * tbabs* (polconst * diskbb + pollin * relxill)}. Consistent with the broadband spectral analysis, the \texttt{nuMLIv1.mod} table model was explicitly included to correct for the \textit{NuSTAR} FPMB MLI tear, ensuring that the instrumental soft-excess did not bias the joint fit. Specifically, the MLI fraction parameter was allowed to vary freely for the FPMB data group, while it was strictly frozen to 1.0 for the FPMA and all three \textit{IXPE} detector units. This approach mathematically isolates the correction to the affected FPMB module without altering the continuum of the unaffected instruments. In this configuration, the \texttt{polconst} component applies a constant polarization to the direct thermal emission from the \texttt{diskbb} disk (which was allowed to vary during the fit to observationally test for intrinsic disk polarization). The linear polarization operator \texttt{pollin} maps the polarization degree of the coronal reflection component (\texttt{relxill}) as an energy-dependent linear function. To avoid artificially underestimating the statistical uncertainties due to parameter correlations \citep{DiMarco2026}, the polarization angles (PAs) for the disk and coronal components were strictly untied and allowed to vary independently.

\section{Result}
\label{sec:results}

\subsection{Evolution of the Outburst}
\label{subsec:outburst_evolution}

\begin{figure}[h!]
    \centering
    \includegraphics[width=0.48\textwidth,height=9cm]{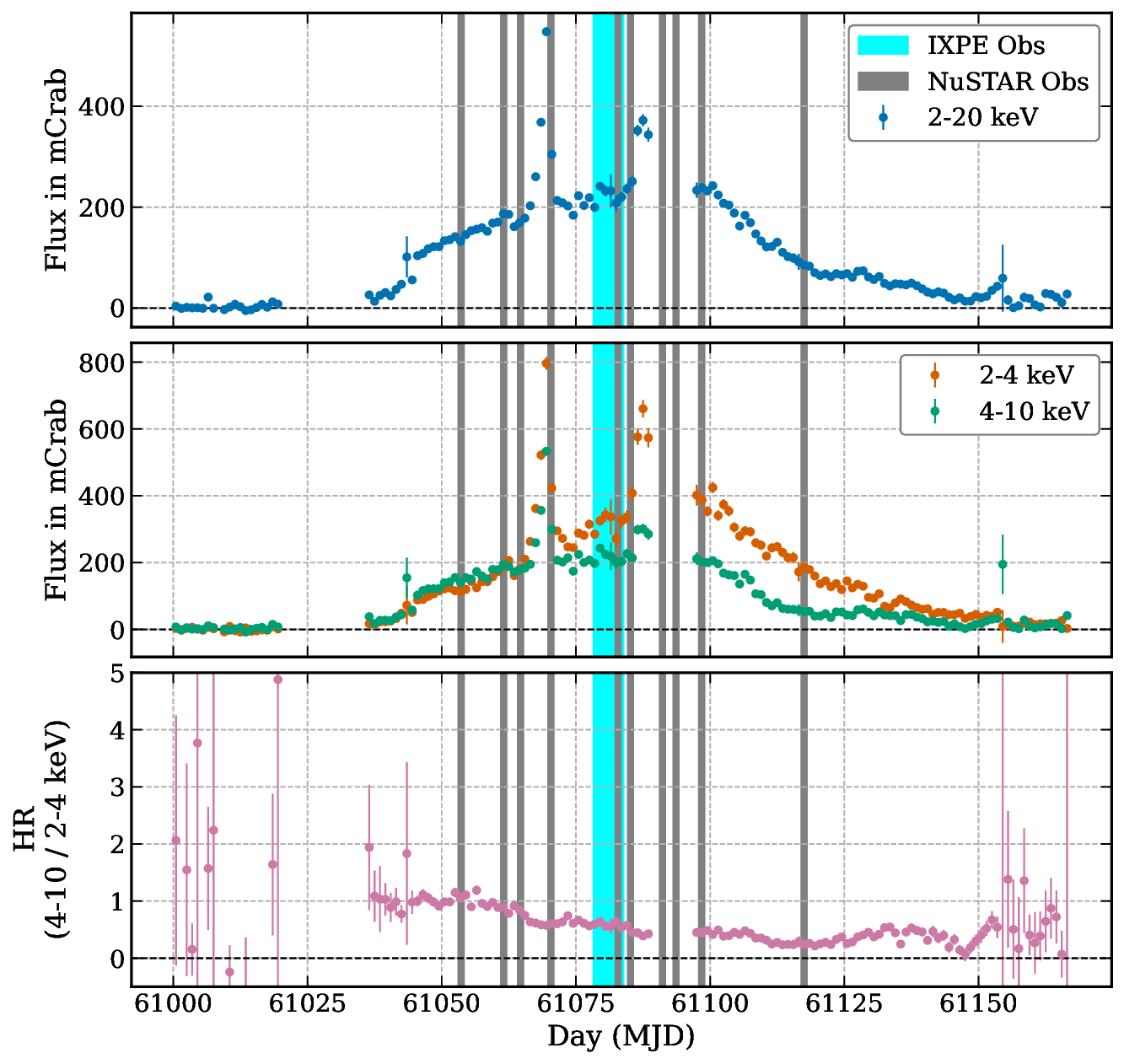}
    \caption{Daily averaged MAXI/GSC light curves of GS 1354-64 during the 2025--2026 outburst. From top to bottom: total $2-20$ keV flux, soft $2-4$ keV flux, hard $4-10$ keV flux (all in mCrab), and the Hardness Ratio (HR, $4-10$ keV / $2-4$ keV). The vertical shaded bands indicate the epochs of the ten \textit{NuSTAR} and the single \textit{IXPE} observations.}
    \label{fig:maxi_lc}
\end{figure}

\begin{figure*}
    \centering
    \includegraphics[width=0.90\textwidth]{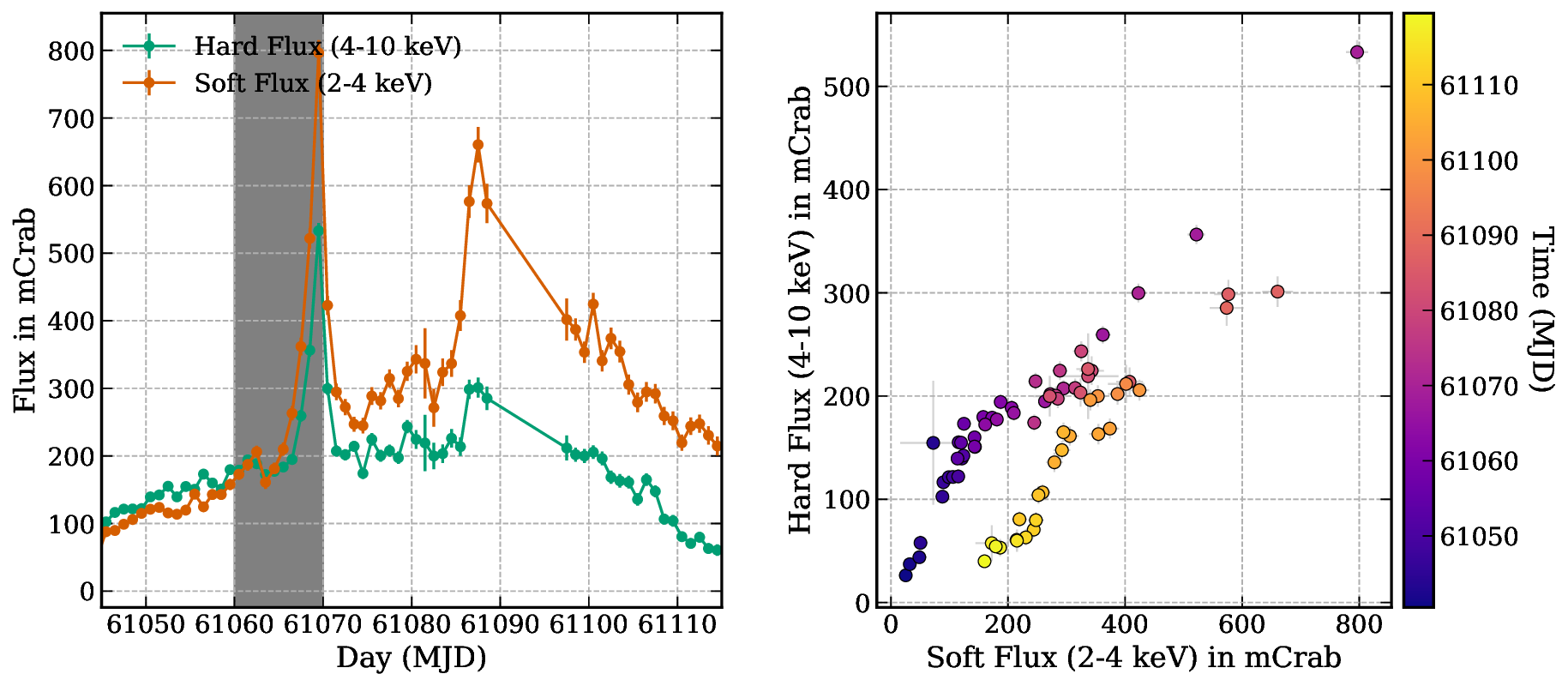}
    \caption{\textit{Left:} Overlaid MAXI/GSC light curves of the soft (2--4 keV, orange) and hard (4--10 keV, green) X-ray bands during the peak of the outburst. The shaded region highlights the state transition where the soft flux surges while the hard flux simultaneously collapses. \textit{Right:} A scatter plot of the hard versus soft flux during the same period, color-coded by time (MJD), demonstrating the strong anti-correlation characteristic of Compton cooling of the corona by the advancing thermal disk.}
    \label{fig:anti_correlation}
\end{figure*}

To contextualize the detailed spectral and temporal analyses, I first examine the long-term, broadband evolution of the 2025--2026 outburst using daily monitoring data from the MAXI/GSC instrument. Figure \ref{fig:maxi_lc} presents the multi-band light curves ($2-20$ keV, $2-4$ keV, and $4-10$ keV) alongside the Hardness Ratio (HR), defined as the ratio of the $4-10$ keV flux to the $2-4$ keV flux. The ten \textit{NuSTAR} and the single \textit{IXPE} observation epochs analyzed in this work are overlaid as shaded vertical bands, demonstrating the comprehensive coverage across all major phases of the outburst.

The outburst commenced with a rapid rise in the hard X-ray flux, characteristic of a canonical hard state. During this initial rising phase (prior to MJD $\sim 61060$), the emission was slightly dominated by the $4-10$ keV band, driving the HR to high values. Following this peak, the system exhibited a sharp spectral softening. The $2-4$ keV flux surged while the $4-10$ keV flux simultaneously declined, causing the HR to plummet. This anti-correlation is the hallmark of a state transition, marking the collapse of the inner corona and the emergence of the geometrically thin, optically thick accretion disk as the dominant emission component in the soft state.

The physical nature of this transition is explicitly visualized in Figure \ref{fig:anti_correlation}. The left panel highlights the temporal crossing of the fluxes, while the right panel presents a flux-flux scatter plot parameterized by time. At the onset of the outburst (indicated by the darkest, early-time markers), both fluxes rise together, tracing a steep path from the bottom-left near the origin up toward the top-left quadrant. As the system reaches the peak of the hard X-ray flux (the apex in the upper-left), the state transition initiates. From this peak, the system dynamically traces a sharp diagonal path downward and to the right, ending in the bottom-right quadrant (characterized by maximal soft flux and suppressed hard flux). This distinct negative slope graphically captures the strong anti-correlation between the two bands during the transition. Physically, this diagonal track is the direct observational signature of Compton cooling: as the thermal accretion disk advances inward, it floods the inner region with soft seed photons. These photons rapidly drain energy from the hot corona via inverse-Compton scattering, thereby suppressing the hard X-ray emission exactly as the soft disk emission peaks.

\begin{figure}[h!]
    \centering
    \includegraphics[width=0.48\textwidth]{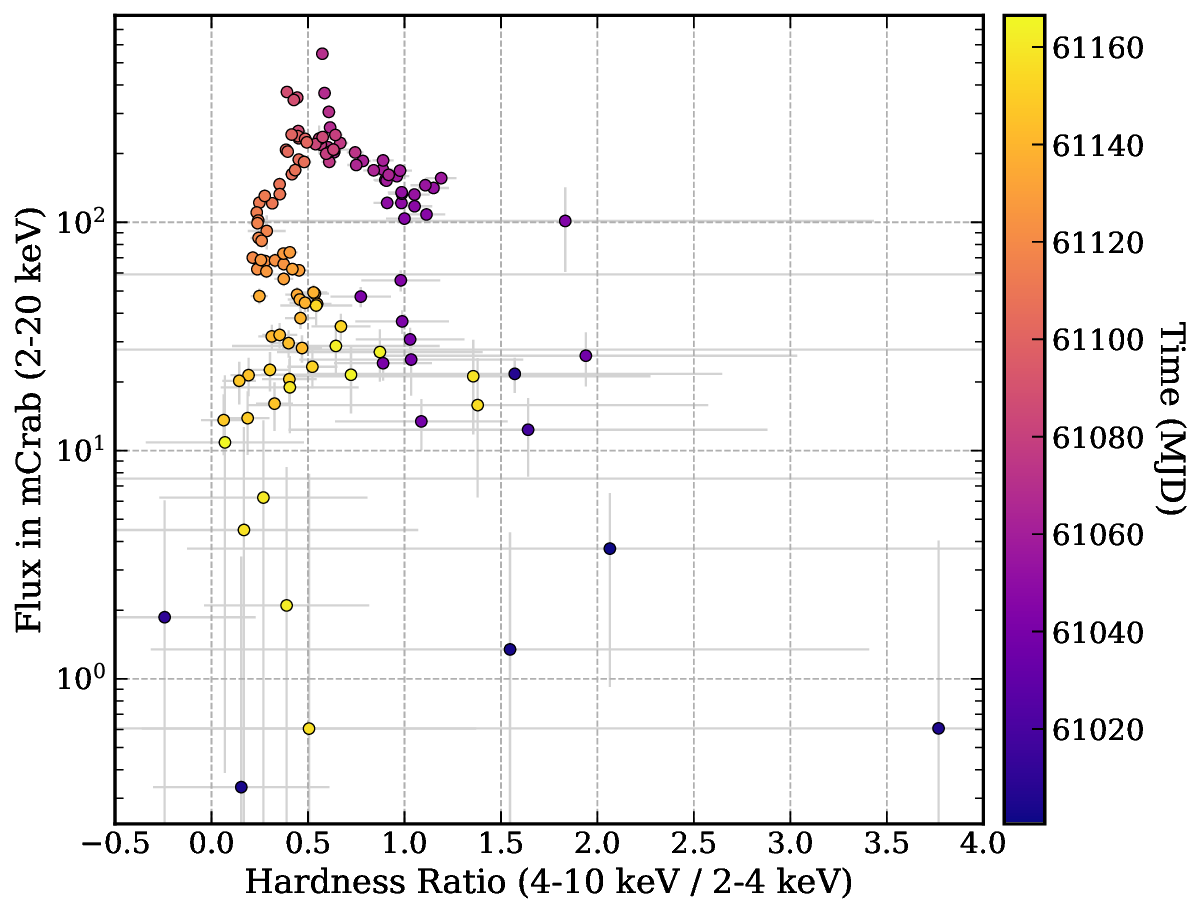}
    \caption{Time-resolved Hardness-Intensity Diagram (HID) of GS 1354-64 using MAXI/GSC data. The color map represents the time evolution in MJD, illustrating the classical counter-clockwise ``q''-shaped track indicative of a complete state transition.}
    \label{fig:maxi_hid}
\end{figure}

The global state transitions of the system are best visualized through the Hardness-Intensity Diagram (HID) shown in Figure \ref{fig:maxi_hid}, where the total $2-20$ keV flux is plotted against the HR. The color-coding tracks the temporal evolution (in MJD) of the outburst. The source traces out a distinct, counter-clockwise ``q''-shaped loop. It begins in the bottom right (hard and faint), ascends vertically as the mass accretion rate increases in the hard state, and then moves horizontally to the left (softening at a near-constant intensity) across the intermediate states. It subsequently drops in intensity while remaining in the soft state, before finally hardening and returning to the bottom right as the outburst decays.

Crucially, this complete ``q''-shaped track confirms that GS 1354-64 successfully completed a full hard-to-soft-to-hard state transition during the 2025--2026 event. This represents a significant departure from its well-documented outbursts in 1997 and 2015, which were classified as ``failed'' outbursts that never left the hard/hard-intermediate states.

\subsection{Spectral Properties}
\label{subsec:spectral_results}

\begin{figure}[ht!]
    \centering
    \includegraphics[width=0.48\textwidth,height=14cm]{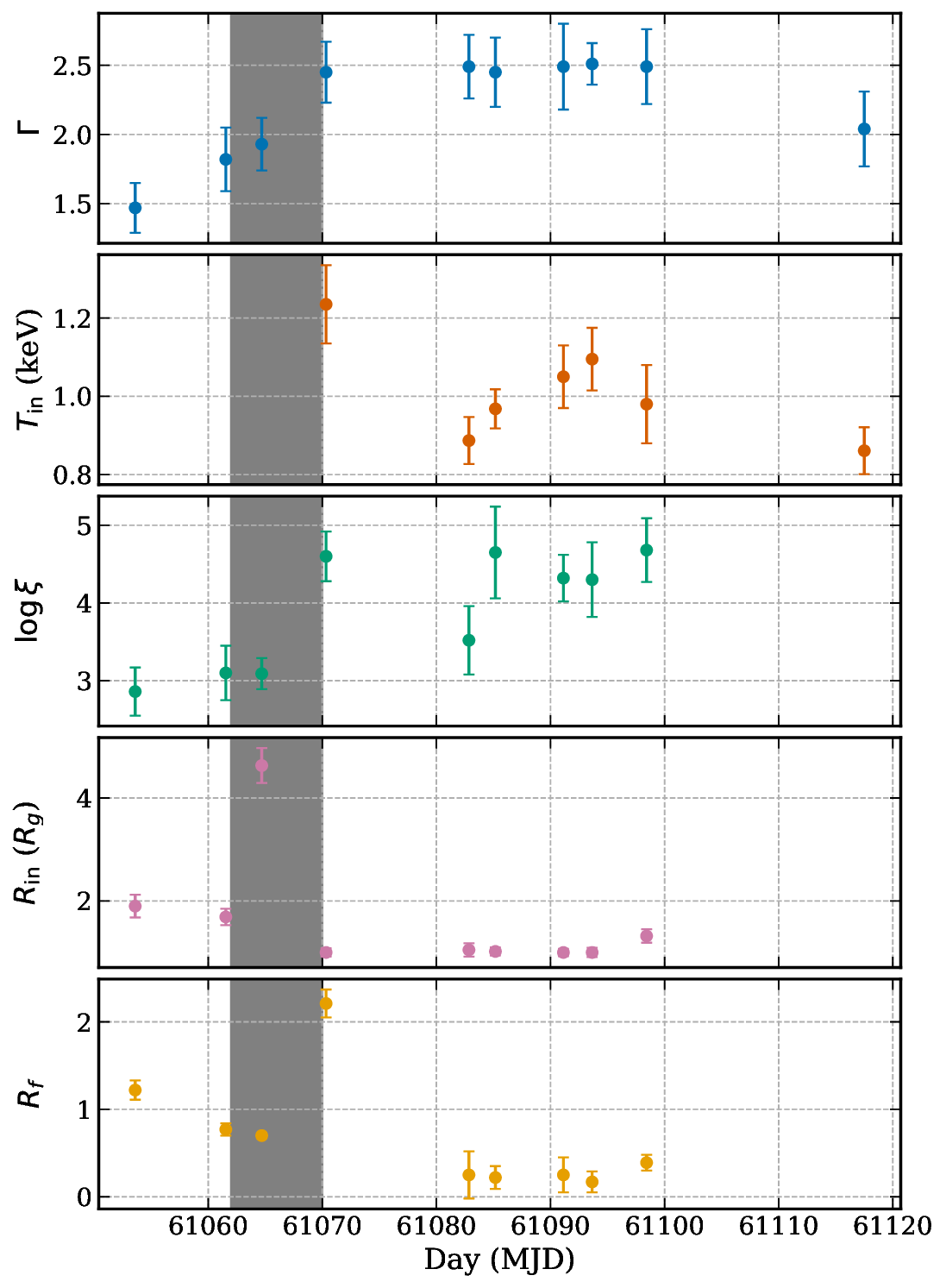}
    \caption{Temporal evolution of the best-fit spectral parameters for GS 1354-64 utilizing the \textit{NuSTAR} data. From top to bottom: the power-law photon index ($\Gamma$), inner disk temperature ($T_{\rm in}$), ionization parameter ($\log \xi$), inner disk radius ($R_{\rm in}$), and reflection fraction ($R_f$). Error bars represent the 90\% confidence intervals. The shaded gray vertical band roughly indicates the rapid state transition phase.}
    \label{fig:spec_evol}
\end{figure}

\begin{table*}[t]
\centering
\caption{Best-Fit Spectral Parameters of GS 1354-64 across the NuSTAR Observations. The foreground absorption ($N_H$) is strictly frozen to $0.70 \times 10^{22}~cm^{-2}$ across all epochs. The thermal disk temperature ($T_{in}$), Gaussian absorption line energy ($E_{line}$), and line width ($\sigma$) are reported in keV \ref{subsec:spectral_results}.}
\label{tab:spectral_pars}
\resizebox{\textwidth}{!}{%
\begin{tabular}{lcccccccccccc}
\hline\hline
ObsID & MJD & $T_{in}$ & $\Gamma$ & $E_{line}$ & $\sigma$ & $a$ & $i~ (^\circ)$ & $R_{in}~(R_g)$ & $\log\xi$ & $A_{Fe}$ & $R_f$ & $\chi^2/DOF$ \\
\hline
91201302002 & 61053.60 & -- & $1.47\pm0.18$ & $7.28\pm0.45$ & $0.25\pm0.03$ & $0.998\pm0.112$ & $52.82\pm6.60$ & $1.90\pm0.22$ & $2.86\pm0.31$ & $0.59\pm0.05$ & $1.22\pm0.11$ & 2458.42/2429 \\
91201302004 & 61061.54 & -- & $1.82\pm0.23$ & $7.33\pm0.76$ & $0.26\pm0.03$ & $0.998\pm0.103$ & $47.67\pm5.51$ & $1.69\pm0.16$ & $3.10\pm0.35$ & $0.50\pm0.06$ & $0.77\pm0.07$ & 2151.80/2057 \\
91201302006 & 61064.69 & -- & $1.93\pm0.19$ & $7.29\pm0.52$ & $0.13\pm0.05$ & $0.998\pm0.100$ & $49.59\pm2.38$ & $4.63\pm0.34$ & $3.09\pm0.20$ & $0.50\pm0.06$ & $0.70\pm0.03$ & 2052.43/1915 \\
91202308002 & 61070.33 & $1.235\pm0.10$ & $2.45\pm0.22$ & -- & -- & $0.998\pm0.091$ & $49.35\pm4.71$ & $1.00\pm0.08$ & $4.60\pm0.32$ & $0.50\pm0.04$ & $2.21\pm0.16$ & 1785.52/1714 \\
91202311002 & 61082.85 & $0.887\pm0.06$ & $2.49\pm0.23$ & -- & -- & $0.953\pm0.099$ & $47.99\pm3.25$ & $1.05\pm0.13$ & $3.52\pm0.44$ & $1.00\pm0.06$ & $0.25\pm0.27$ & 1412.34/1479 \\
91202311004 & 61085.17 & $0.968\pm0.05$ & $2.45\pm0.25$ & -- & -- & $0.970\pm0.098$ & $44.75\pm3.50$ & $1.02\pm0.08$ & $4.65\pm0.59$ & $4.64\pm0.18$ & $0.22\pm0.13$ & 1352.40/1377 \\
91202311006 & 61091.13 & $1.050\pm0.08$ & $2.49\pm0.31$ & -- & -- & $0.960\pm0.090$ & $51.71\pm3.66$ & $1.00\pm0.07$ & $4.32\pm0.30$ & $2.22\pm0.10$ & $0.25\pm0.20$ & 1281.43/1349 \\
91202311008 & 61093.65 & $1.095\pm0.08$ & $2.51\pm0.15$ & -- & -- & $0.947\pm0.076$ & $51.18\pm3.93$ & $1.00\pm0.09$ & $4.30\pm0.48$ & $2.68\pm0.15$ & $0.17\pm0.12$ & 1403.53/1381 \\
91202311010 & 61098.42 & $0.980\pm0.10$ & $2.49\pm0.27$ & -- & -- & $0.998\pm0.127$ & $51.79\pm4.77$ & $1.32\pm0.13$ & $4.68\pm0.41$ & $4.93\pm0.08$ & $0.39\pm0.09$ & 1331.23/1356 \\
91202311012 & 61117.50 & $0.861\pm0.06$ & $2.04\pm0.27$ & -- & -- & -- & -- & -- & -- & -- & -- & 719.54/655 \\
\hline
\end{tabular}%
}
\end{table*}

The broadband X-ray spectral evolution of the 2025--2026 outburst from GS 1354-64 was rigorously tracked across ten \textit{NuSTAR} observations spanning from MJD 61053 to 61117. To systematically capture the underlying physical changes while avoiding instrumental artifacts, I applied a combined phenomenological and relativistic reflection modeling framework within \texttt{XSPEC}, with the line-of-sight absorption strictly frozen to $N_H = 0.70 \times 10^{22}\ \rm cm^{-2}$ after applying the \texttt{nuMLIv1.mod} correction. Depending on the outburst state, the continuum was modeled using a multi-color disk blackbody (\texttt{diskbb}) and either a relativistic reflection component (\texttt{relxill}) or a standard power-law (\texttt{powerlaw}). The complete set of best-fit spectral parameters, including estimated uncertainties, is presented in Table \ref{tab:spectral_pars}.

To better visualize the physical changes driving these state transitions, I present the temporal evolution of the primary spectral parameters across the ten \textit{NuSTAR} epochs in Figure \ref{fig:spec_evol}. The top panel illustrates the softening of the primary continuum, with the photon index ($\Gamma$) rising sharply from the initial hard state ($\sim 1.47$) to plateau during the soft state ($\sim 2.50$), before dropping again as the outburst decays. Concurrently, the emergence and stabilization of the accretion disk temperature ($T_{\rm in} \sim 1$ keV) is clearly visible. Because the instrumental soft-excess was properly corrected, the geometric and environmental evolution of the inner accretion flow is now robustly and cleanly captured: the inner disk radius ($R_{\rm in}$) collapses toward the ISCO ($< 2 R_g$) during the soft state, accompanied by a dramatic increase in the ionization parameter ($\log \xi$). Meanwhile, the reflection fraction ($R_f$) exhibits a sharp spike during the initial transition before dropping to suppressed levels ($R_f \sim 0.2 - 0.4$) for the remainder of the luminous soft phase.

\textbf{The Hard State and Highly Ionized Disk Winds:}
During the initial rising phase of the outburst (MJD 61053 to 61061), the source resided in a definitive hard state. The broadband continuum was entirely dominated by strong Comptonization from the hot corona, characterized by a hard photon index ($\Gamma = 1.47 \pm 0.18$ to $1.82 \pm 0.23$). In this regime, the thermal emission from the inner accretion disk was either too cool or the disk was too truncated to be detected within the \textit{NuSTAR} bandpass. Notably, significant absorption residuals were detected in the Fe K band at $\sim 7.3$ keV. The inclusion of a Gaussian absorption line model (\texttt{gabs}) effectively resolved these residuals, providing strong evidence for the presence of highly ionized equatorial disk winds---likely associated with Fe XXVI---exhibiting an equivalent line width of $\sigma = 0.13 \pm 0.05$ to $0.26 \pm 0.03$ keV.

\textbf{The Hard-to-Soft State Transition:}
By MJD 61064, the source initiated a distinct state transition, passing through a hard-intermediate state where the primary power-law emission softened to $\Gamma = 1.93 \pm 0.19$. Shortly thereafter, beginning around MJD 61070, the accretion dynamics fundamentally shifted as the source entered the canonical soft state. During this phase, the multi-color disk blackbody (\texttt{diskbb}) component became statistically necessary and physically dominant at the lower end of the \textit{NuSTAR} passband. This revealed a hot, optically thick inner accretion disk with temperatures stabilizing between $T_{\rm in} = 0.88 \pm 0.06$ to $1.23 \pm 0.10$ keV. Concurrently, the Comptonized corona cooled and softened significantly, with the photon index settling into a tight distribution of $\Gamma = 2.45 \pm 0.25$ to $2.51 \pm 0.15$.

\textbf{Relativistic Reflection and System Geometry:}
Throughout the active reflection-dominated states (MJD 61053 to 61098), the relativistically broadened Fe K$\alpha$ line and Compton hump allowed us to constrain the inner accretion geometry. Because fitting these observations individually left some geometric parameters loosely constrained, we applied a joint spectral fit across all these epochs (see Figure \ref{fig:joint_fit_spec}). By tying the static system parameters together, the joint fit achieved a statistically excellent global minimum ($\chi^2_\nu = 1.09$). This joint modeling securely anchors the inner disk inclination to $i = 51.55^\circ \pm 0.17^\circ$ and yields a highly stable sub-solar iron abundance of $A_{\rm Fe} = 0.620 \pm 0.007$. Furthermore, the black hole spin is consistently driven to the maximal theoretical limit ($a_* \approx 0.998$). This definitively confirms that GS 1354-64 harbors a rapidly rotating black hole, and during the luminous soft state, its accretion disk rigidly pins to the ISCO.

\textbf{Outburst Decay and the Fading of Reflection:}
In the final \textit{NuSTAR} observation (MJD 61117), obtained approximately 20 days after the preceding exposure, the overall X-ray flux dropped substantially as the source exhibited distinct characteristics of outburst decay. During this faint phase, the relativistic reflection features faded entirely into the background noise, rendering the \texttt{relxill} model statistically unnecessary and prone to artificial degeneracies. By replacing the reflection component with a simple \texttt{powerlaw}, the continuum was successfully and physically modeled. The overall continuum hardened significantly ($\Gamma = 2.04 \pm 0.27$), and the thermal disk cooled to $T_{\rm in} = 0.86 \pm 0.06$ keV. The application of this simpler, physically motivated model naturally resolves the unphysical parameter swings that can occur when over-parameterizing low-luminosity data, unambiguously confirming the weakening of the continuous mass accretion rate as GS 1354-64 returned to quiescence.

\subsection{Timing Properties}
\label{subsec:temporal_results}

\begin{figure*}
    \centering
    \begin{subfigure}[b]{0.33\textwidth}
        \centering
        \includegraphics[width=\textwidth]{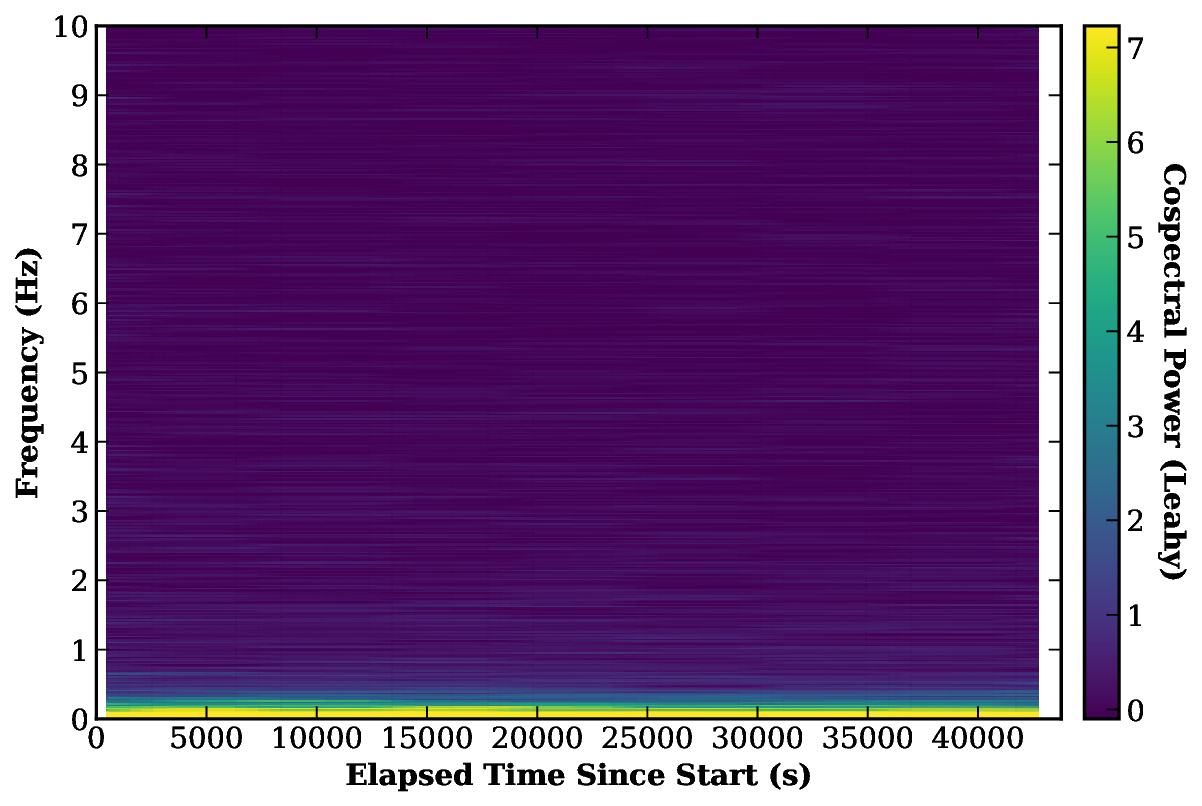}
    \end{subfigure}\hfill
    \begin{subfigure}[b]{0.33\textwidth}
        \centering
        \includegraphics[width=\textwidth]{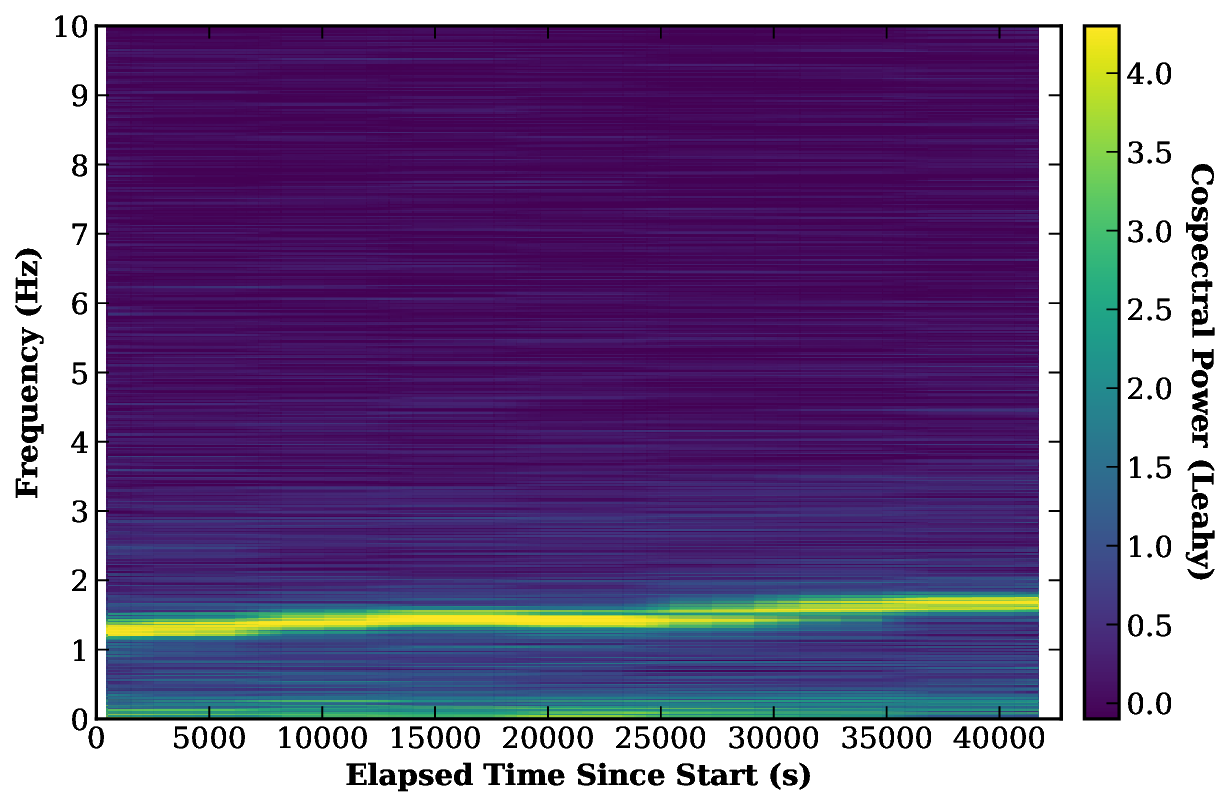}
    \end{subfigure}\hfill
    \begin{subfigure}[b]{0.33\textwidth}
        \centering
        \includegraphics[width=\textwidth]{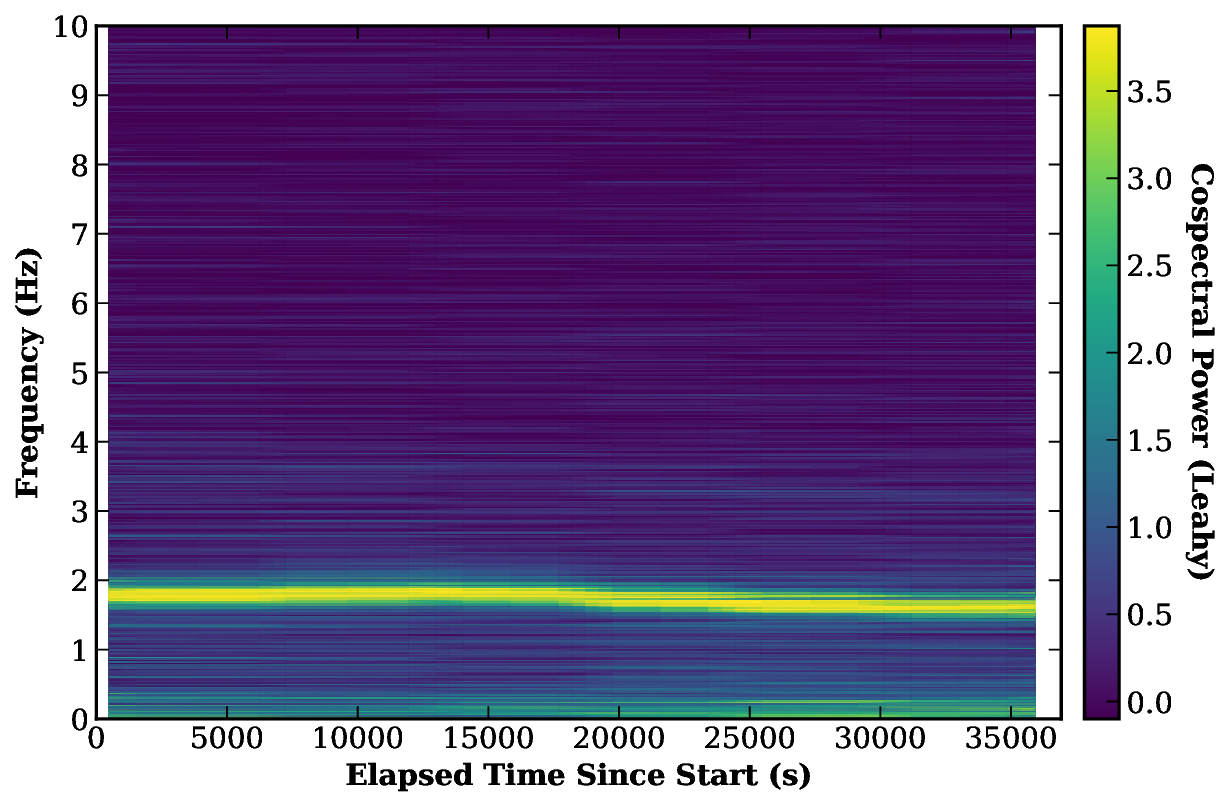}
    \end{subfigure}
    \caption{\textit{NuSTAR} dynamic cospectra tracking the structural evolution of the Type-C QPO during the initial rising hard state of GS 1354--64. \textbf{Left:} MJD 61053.60 (ObsID 91201302002) showing a highly coherent, low-frequency feature at $\sim 0.08$ Hz. \textbf{Center:} MJD 61061.54 (ObsID 91201302004) capturing the QPO track dynamically migrating to $\sim 1.4-1.6$ Hz. \textbf{Right:} MJD 61064.69 (ObsID 91201302006) showing the QPO peaking at $\sim 1.73$ Hz immediately prior to the state transition. The color scale represents Leahy-normalized cospectral power, utilizing 1024-second continuous data segments to safely bypass instrumental dead-time artifacts while maintaining high signal-to-noise. This visible upward drift in centroid frequency observationally traces the geometric compression of the precessing hot inner flow.}
    \label{fig:dynamic_qpo_evolution}
\end{figure*}

Before examining the detailed time-averaged fits, the dynamic evolution of the accretion geometry is strikingly visible in the dynamic cospectra. As shown in Figure \ref{fig:dynamic_qpo_evolution}, during the initial rising hard state, the source exhibits bright, highly coherent Type-C QPO tracks that are easily resolvable above the Poisson noise floor within single 1024-second orbital segments. Across these first three epochs, the oscillation is observed to systematically migrate from $\sim0.08$ Hz up to $\sim1.73$ Hz. This distinct, unbroken upward drift provides direct visual evidence for a rapid, macroscopic evolution of the inner accretion geometry. Conversely, during the subsequent soft-state observations (MJD $\ge$ 61070), the dynamical cospectra visually appear featureless. This is not an instrumental artifact, but a direct physical consequence of the state transition. In a power density spectrum, the integrated power of a QPO scales proportionally to the square of its fractional RMS amplitude ($Power \propto RMS^2$). As the massive hard-state corona collapses, the fractional RMS of the QPO plummets from $\sim 10\%$ to $\sim 3\%$, effectively dropping the Fourier power by an order of magnitude. Furthermore, the soft-state QPOs become broader, smearing this severely reduced power across a wider range of frequency bins. Consequently, the peak amplitude of the signal in any single 1024-second snapshot falls completely below the random statistical variance (the Poisson noise floor) inherent to short exposures. Therefore, to extract the faint, high-frequency boundary oscillations residing in the soft state, it is mathematically required to rely exclusively on the full time-averaged cospectra. By averaging dozens of 1024-second segments across the entire observation, the random Poisson fluctuations are sufficiently suppressed, allowing the residual soft-state QPOs to emerge with high statistical significance (for visual confirmation of these dynamic detection limits, see Appendix \ref{sec:appendix_soft_state} and Figure \ref{fig:soft_state_dynamic}).

To track the fast timing evolution of the black hole candidate GS 1354-64 during its 2025--2026 outburst, I conducted a systematic Fourier timing analysis across all ten NuSTAR epochs. We extracted deadtime-corrected cospectra from co-aligned modules FPMA and FPMB in the 3.0--79.0~keV energy range using a temporal resolution of $\Delta t = 0.005$~s, providing a Nyquist frequency of 100~Hz. The dynamical cross-spectra (cospectra) were computed to isolate coherent astronomical variability from uncorrelated instrumental Poisson noise.

\begin{figure}[ht!]
    \centering
    \includegraphics[width=\columnwidth,height=9.5cm]{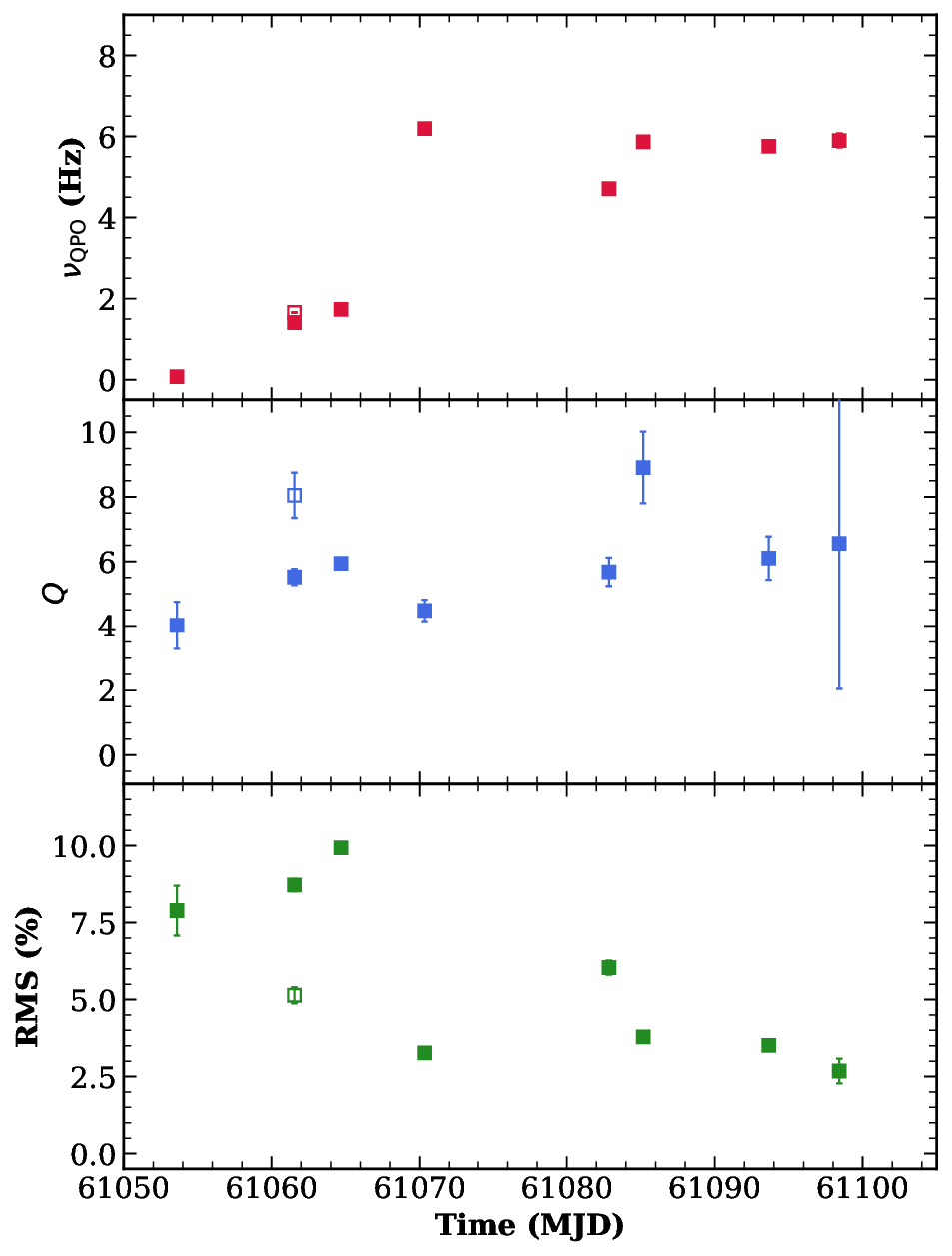}
    \caption{Evolution of the low-frequency Quasi-Periodic Oscillation (LFQPO) timing properties for GS 1354-64 as a function of time (MJD) across the 2025--2026 outburst. Panels from top to bottom illustrate the standalone variations of the QPO centroid frequency $\nu_{\mathrm{QPO}}$ (crimson), quality factor $Q$ (royalblue), and background-corrected fractional RMS amplitude (forestgreen). Filled square markers signify the fundamental tracks, while open square markers represent the secondary upper components resolved during the multi-component state.}
    \label{fig:qpo_evolution}
\end{figure}

The power density spectra reveal a rich collection of low-frequency Quasi-Periodic Oscillations (LFQPOs) whose physical properties show a dramatic dependence on the outburst evolution stage. The chronological tracking of the QPO centroid frequency ($\nu_{\mathrm{QPO}}$), quality factor ($Q = \nu_0/\Delta\nu$), and background-dilution corrected fractional root-mean-square amplitude ($\mathrm{RMS}$) is displayed visually in Figure~\ref{fig:qpo_evolution}, while the exact fitted values and corresponding statistical uncertainties are compiled comprehensively in Table~\ref{tab:qpo_parameters}.

In the initial low-hard state (LHS) epoch (ObsID 91201302002; MJD 61053.60), a weak but coherent low-frequency feature is detected at $\nu_{\mathrm{QPO}} = 0.08 \pm 0.01$~Hz with a quality factor $Q = 4.02 \pm 0.73$ and a notable fractional intensity ($\mathrm{RMS} = 7.89\% \pm 0.81\%$). As the source expanded its mass accretion rate and progressed along the rising phase of the outburst toward the hard-intermediate state (HIMS; MJD 61061.54, ObsID 91201302004), the timing structure became significantly more complex. During this epoch, the spectrum cleanly resolves a dual-peaked structure: a fundamental QPO at $\nu_1 = 1.41 \pm 0.00$~Hz ($Q = 5.52 \pm 0.25$, $\mathrm{RMS} = 8.72\% \pm 0.20\%$) alongside a prominent, highly coherent upper component at $\nu_2 = 1.66 \pm 0.01$~Hz ($Q = 8.05 \pm 0.70$, $\mathrm{RMS} = 5.14\% \pm 0.26\%$). By the subsequent observation on MJD 61064.69 (ObsID 91201302006), these components evolved into a singular, intense LFQPO at $\nu_{\mathrm{QPO}} = 1.73 \pm 0.00$~Hz, reaching its maximum modulation strength ($\mathrm{RMS} = 9.93\% \pm 0.11\%$, $Q = 5.94 \pm 0.16$).

Following a data gap corresponding to the peak of the outburst and early decay, the source re-stabilized into a significantly higher frequency domain ($\nu_{\rm QPO} \approx 4.71 - 6.20$ Hz), as reported in Table 3. Interestingly, on MJD 61082.85 (ObsID 91202311002), the QPO is distinctly detected at $\nu_{\rm QPO} = 4.71 \pm 0.02$ Hz with a strong recovered coherence ($Q = 5.68 \pm 0.44$) and a fractional RMS of $6.04\% \pm 0.23\%$, indicating a stable localized oscillating boundary before further evolution. The QPOs shift to a significantly higher frequency domain ($4.71 - 6.20$ Hz) with variable coherence. The fractional RMS drops to $\sim2.6 - 6.0\%$, and the features exhibit fluctuating widths.

\begin{table}[ht!]
\scriptsize
\centering
\caption{Evolution of the Low-Frequency Quasi-Periodic Oscillation (LFQPO) Parameters for GS 1354-64 Derived from NuSTAR Observations.}
\label{tab:qpo_parameters}
\begin{tabular}{lccccc}
\toprule
Observation ID & Average MJD & $\nu_{\mathrm{QPO}}$ (Hz) & Quality Factor ($Q$) & RMS (\%) \\
\midrule
91201302002 & 61053.60 & $0.08 \pm 0.01$ & $4.02 \pm 0.73$ & $7.89 \pm 0.81$ \\
91201302004 & 61061.54 & $1.41 \pm 0.01$ & $5.52 \pm 0.25$ & $8.72 \pm 0.20$ \\
            &          & $1.66 \pm 0.01$ & $8.05 \pm 0.70$ & $5.14 \pm 0.26$ \\
91201302006 & 61064.69 & $1.73 \pm 0.01$ & $5.94 \pm 0.16$ & $9.93 \pm 0.11$ \\
91202308002 & 61070.33 & $6.20 \pm 0.03$ & $4.48 \pm 0.33$ & $3.27 \pm 0.10$ \\
91202311002 & 61082.85 & $4.71 \pm 0.02$ & $5.68 \pm 0.44$ & $6.04 \pm 0.23$ \\
91202311004 & 61085.17 & $5.87 \pm 0.03$ & $8.91 \pm 1.11$ & $3.79 \pm 0.18$ \\
91202311006 & 61091.13 & \multicolumn{3}{c}{--- No QPO Detected ---} & \\
91202311008 & 61093.65 & $5.76 \pm 0.04$ & $6.10 \pm 0.67$ & $3.51 \pm 0.13$ \\
91202311010 & 61098.42 & $5.90 \pm 0.17$ & $6.56 \pm 4.51$ & $2.68 \pm 0.40$ \\
91202311012 & 61117.50 & \multicolumn{3}{c}{--- No QPO Detected ---} & \\
\bottomrule
\end{tabular}
\end{table}

\subsection{Polarization Properties}
\label{subsec:polarimetry_results}

To constrain the geometry of the scattering media, I present both model-independent and model-dependent polarimetric results, unified in Table \ref{tab:combined_polarization}. The joint broadband spectro-polarimetric modeling of the simultaneous soft-state data window (MJD 61082.85), as shown in Figure \ref{fig:joint_2col_spec} and detailed in Table \ref{tab:joint_fit_table}, incorporating all three IXPE detector units (DU1, DU2, DU3) and un-subtracted background spectra, achieved an excellent fit quality of $\chi^2/d.o.f.=3013.05/2803$ ($\chi_v^2 \approx 1.07$). To break the mathematical degeneracy in the soft X-ray band, the line-of-sight column density was fixed to the value of $N_H = 7.0 \times 10^{21}\ \rm cm^{-2}$ \citep{Koljonen2016}, a standard gain correction was applied to the \textit{IXPE} Stokes $I$ spectra, and the FPMB MLI fraction was allowed to vary to properly account for the thermal blanket tear.

\begin{figure}[ht!]
    \centering
    \includegraphics[width=0.85\columnwidth,height=9cm]{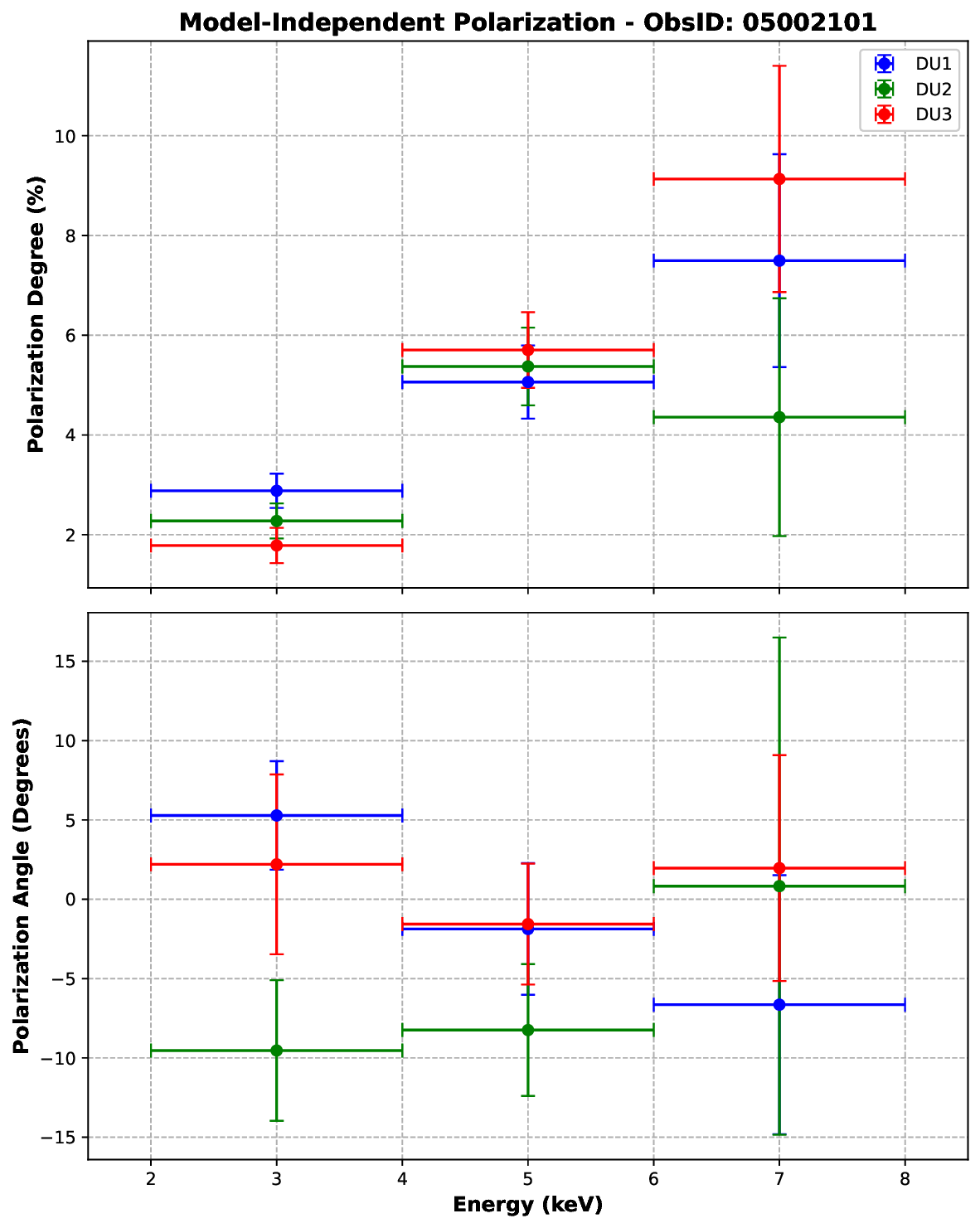}
    \caption{Energy-dependent Polarization Degree (top) and Polarization Angle (bottom) for GS 1354-64 during the IXPE observation (MJD 61078--61083). Data from DU1 (blue), DU2 (green), and DU3 (red) are shown. A strong increase in polarization degree is observed at higher energies, while the polarization angle remains consistently distributed near $0^\circ$.}
    \label{fig:pol_energy}
\end{figure}

\begin{figure}
    \centering
    \includegraphics[width=0.85\columnwidth,height=9cm]{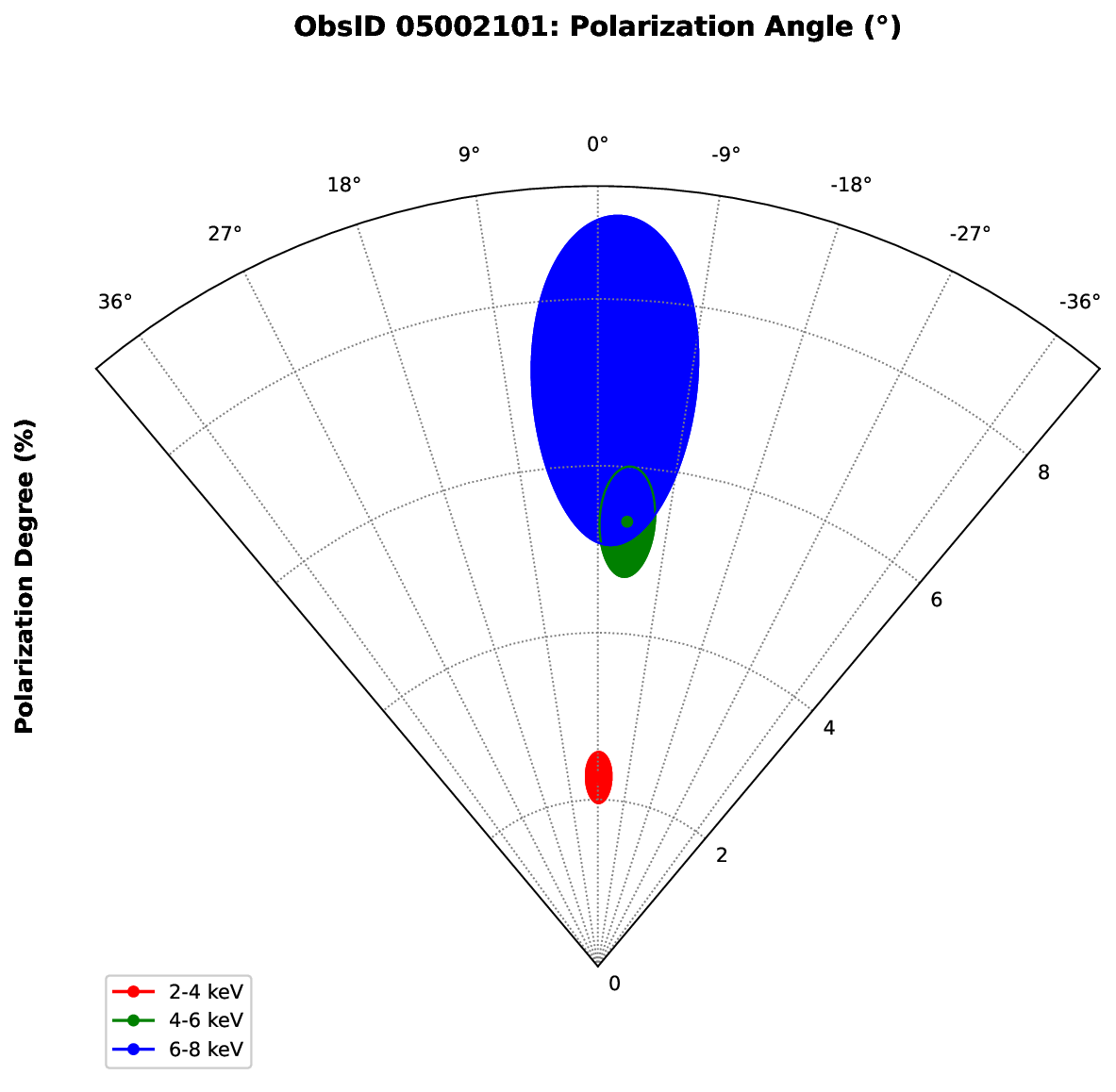}
    \caption{Radial confidence contours of the Polarization Degree versus Polarization Angle for the narrow energy bins: 2--4 keV (red), 4--6 keV (green), and 6--8 keV (blue). The radial distance represents the polarization percentage, while the angular coordinate represents the polarization angle. Contours represent the 68\%, 90\%, and 99\% confidence levels, illustrating the strict, energy-independent alignment of the polarization angle.}
    \label{fig:pol_contour}
\end{figure}

During the fitting procedure, I explicitly allowed the intrinsic polarization of the thermal disk to vary. However, the fit statistically constrained the disk polarization to an upper limit of $< 0.1\%$, observationally proving that the thermal disk is inherently unpolarized in this energy band. Consequently, the independent polarization angle of the thermal disk is mathematically undefined and completely unconstrained in the fit. In stark contrast, the coronal component displays a highly significant energy-increasing polarization degree. The intrinsic coronal polarization climbs steeply with a robust positive slope of $A_{\rm slope} = 1.3\% \pm 0.2\%$ per keV, tracking perfectly with the upper bounds measured via the model-independent PCUBE channels. Furthermore, while left completely free to vary during this localized spectro-polarimetric modeling, the reflection inclination ($i=54.9^\circ\pm5.0^\circ$) and iron abundance ($A_{\rm Fe}=1.04\pm0.65$) agrees well with of the rigid global joint-fit values ($i=51.55^\circ$, $A_{\rm Fe}=0.620$). This independent convergence acts as a powerful self-consistency check, proving that the system's fundamental geometry remains completely stable and naturally anchors the physical interpretation of the polarimetric vectors.

An independent analysis of the identical \textit{IXPE} and \textit{NuSTAR} data by \citet{Ravi2026} reported similar polarimetric properties. My primary results strongly agree with their findings, specifically the detection of a highly significant, energy-increasing polarization degree with a stable polarization angle \citep{Ravi2026}. However, the present analysis introduces key methodological advancements over this prior work. While \citet{Ravi2026} excluded \textit{IXPE} DU2 and relied on phenomenological reflection modeling (\texttt{nthComp} + \texttt{smedge} + \texttt{laor}) with an assumed unpolarized disk, I successfully integrated all three \textit{IXPE} detector units. Furthermore, employing the physically self-consistent \texttt{relxill} model robustly constrained the system's inclination, and the spectral fit statistically proved the thermal disk is intrinsically unpolarized rather than relying on an \textit{a priori} assumption.

This massive swing in polarization degree observed via the model-independent channels is illustrated in the energy-dependent polarization profile (Figure \ref{fig:pol_energy}). In the softest band (2--4 keV), dominated by the thermal disk, the polarization remains relatively low ($1.8\% - 2.9\%$). As energy increases into the corona-dominated regime, the polarization climbs to $PD = 5.1\% - 5.7\%$ (4--6 keV) and peaks exceptionally at $\sim 7.5\% - 9.1\%$ in the hard 6--8 keV band. Despite this swing in polarization degree, the polarization angle remains remarkably stable across all energy bins. Fitting a constant model to the energy-binned PA data yields a value of $-2^\circ \pm 1^\circ$, quantitatively confirming the lack of energy-dependent rotation. This stable alignment is visualized in Figure \ref{fig:pol_contour}.

\begin{table}[ht!]
\scriptsize
\centering
\caption{Comprehensive Summary of X-ray Polarization Properties for GS 1354-64. \textit{Top:} Model-independent results derived using the unweighted PCUBE algorithm across broad and narrow energy bins. \textit{Bottom:} Best-fit parameters from the joint spectro-polarimetric modeling (\textit{NuSTAR} + \textit{IXPE}) in \texttt{XSPEC}. Errors on polarization vectors represent robust 90\% confidence intervals.}
\label{tab:combined_polarization}
\renewcommand{\arraystretch}{1.3}
\begin{tabular}{lccc}
\hline\hline
\multicolumn{4}{c}{\textbf{Model-Independent (PCUBE Algorithm)}} \\
\hline
Detector & Energy Band & PD (\%) & PA ($^\circ$) \\
\hline
\multirow{4}{*}{DU1} & 2--8 keV & 3.6 $\pm$ 0.4 & \ \ 2 $\pm$ 3 \\
                     & 2--4 keV & 2.9 $\pm$ 0.3 & \ \ 5 $\pm$ 3 \\
                     & 4--6 keV & 5.1 $\pm$ 0.7 & -2 $\pm$ 4 \\
                     & 6--8 keV & 7.5 $\pm$ 2.1 & -7 $\pm$ 8 \\
\hline
\multirow{4}{*}{DU2} & 2--8 keV & 3.0 $\pm$ 0.4 & -8 $\pm$ 4 \\
                     & 2--4 keV & 2.3 $\pm$ 0.4 & -10 $\pm$ 4 \\
                     & 4--6 keV & 5.4 $\pm$ 0.8 & -8 $\pm$ 4 \\
                     & 6--8 keV & 4.4 $\pm$ 2.4 & \ \ 1 $\pm$ 16 \\
\hline
\multirow{4}{*}{DU3} & 2--8 keV & 3.0 $\pm$ 0.4 & \ \ 1 $\pm$ 4 \\
                     & 2--4 keV & 1.8 $\pm$ 0.4 & \ \ 2 $\pm$ 6 \\
                     & 4--6 keV & 5.7 $\pm$ 0.8 & -2 $\pm$ 4 \\
                     & 6--8 keV & 9.1 $\pm$ 2.3 & \ \ 2 $\pm$ 7 \\
\hline\hline
\multicolumn{4}{c}{\textbf{Model-Dependent (Joint Spectro-Polarimetric Fit)}} \\
\hline
Component & Parameter & \multicolumn{2}{c}{Value} \\
\hline
\texttt{polconst}& Disk PD ($PD_{\rm disk}$) & \multicolumn{2}{c}{$< 0.1\%$ (Pegged)} \\
\texttt{pollin}  & Corona $A_1$ (at 1 keV) & \multicolumn{2}{c}{$1.7\% \pm 0.6\%$} \\
                 & Corona $A_{\rm slope}$ & \multicolumn{2}{c}{$1.3\% \pm 0.2\% \rm\ keV^{-1}$} \\
\texttt{polconst} & Disk PA ($\psi_{\rm disk}$) & Unconstrained & deg \\
\texttt{pollin}   & Corona PA ($\psi_{\rm corona}$) & $-2 \pm 1$ & deg \\
\hline
\end{tabular}
\end{table}

\begin{table}[ht!]
\scriptsize
\centering
\caption{Best-Fit Parameter Matrix from the Joint Spectro-Polarimetric Modeling (\textit{NuSTAR} + \textit{IXPE}), excluding the intrinsic polarization vectors. The corresponding model-dependent polarization parameters for this fit are detailed in Table 4. The foreground absorption ($N_{H}$) was fixed to the value of $7\times10^{22}\rm~cm^{-2}$. Errors on all parameters represent robust 90\% confidence intervals ($\Delta\chi^{2}=2.706$).}
\label{tab:joint_fit_table}
\renewcommand{\arraystretch}{1.2}
\begin{tabular}{llcc}
\hline\hline
Component & Parameter & Best-fit Value & Unit \\
\hline
\texttt{tbabs}   & $N_H$ & $0.70$ (Fixed) & $10^{22}\ \rm cm^{-2}$ \\
\texttt{diskbb}  & $T_{\rm in}$ & $0.98 \pm 0.04$ & $\rm keV$ \\
                 & Normalization & $82.1 \pm 17.9$ & -- \\
\texttt{relxill} & $\Gamma$ & $2.36 \pm 0.02$ & -- \\
                 & $\log\xi$ & $4.70 \pm 0.21$ & $\rm erg\ cm\ s^{-1}$ \\
                 & $A_{\rm Fe}$ & $1.04 \pm 0.65$ & $\rm Solar$ \\
                 & Inclination ($i$) & $54.9 \pm 5.0$ & $\rm deg$ \\
                 & $R_{\rm in}$ & $< 1.05$ (Pegged) & $R_g$ \\
                 & Reflection Fraction ($R_f$) & $2.78 \pm 1.13$ & -- \\
                 & $E_{\rm cut}$ & $300$ (Fixed) & $\rm keV$ \\
                 & Spin ($a$) & $0.998$ (Fixed) & -- \\
\hline
\multicolumn{4}{l}{\textbf{Cross-Calibration Scales}} \\
\hline
\texttt{constant}& $C_{\rm FPMA}$ & $1.00$ (Fixed) & -- \\
                 & $C_{\rm FPMB}$ & $1.01 \pm 0.01$ & -- \\
                 & $C_{\rm DU1}$ & $0.83 \pm 0.01$ & -- \\
                 & $C_{\rm DU2}$ & $0.77 \pm 0.01$ & -- \\
                 & $C_{\rm DU3}$ & $0.72 \pm 0.01$ & -- \\
\hline
\multicolumn{4}{l}{\textbf{Thermal Blanket (MLI) Correction}} \\
\hline
\texttt{mtable}  & $f_{\rm MLI, FPMA}$ & $1.00$ (Fixed) & -- \\
                 & $f_{\rm MLI, FPMB}$ & $0.93 \pm 0.02$ & -- \\
                 & $f_{\rm MLI, IXPE}$ & $1.00$ (Fixed) & -- \\
\hline
\multicolumn{4}{l}{\textbf{Fit Quality}} \\
\hline
                 & Total $\chi^2$ / d.o.f. & $3013.05 / 2803$ & -- \\
\hline\hline
\end{tabular}
\end{table}

\section{Discussion}
\label{sec:discussion}

The \textit{NuSTAR} observations of the 2025--2026 outburst of the black hole candidate GS 1354-64 provide a high-fidelity track of the accretion flow dynamics across multiple spectral states. By employing a combination of thermal disk and relativistic reflection modeling, I have treid to constrain the evolution of the corona, the inner accretion disk geometry, and the presence of ionized outflows.

\subsection{Evolution of the Accretion Flow}
\label{subsec:accretion_evolution}
The outburst followed a standard evolutionary track, originating in a hard, Compton-dominated state before transitioning to a thermal-dominated soft state. During the initial rising phase (MJD 61053 -- 61061), the spectrum was characterized by a hard photon index ($\Gamma = 1.47 \pm 0.18$ to $1.82 \pm 0.23$), indicating that the primary emission was driven by inverse-Compton scattering of soft seed photons within a hot, optically thin corona. The absence of a strong \texttt{diskbb} component in the \textit{NuSTAR} bandpass during this phase suggests that the inner accretion disk was relatively cool, and potentially truncated at a larger radius, consistent with the standard truncated disk paradigm for black hole hard states \citep[e.g.,][]{Done2007}.

As the outburst progressed toward the soft state (MJD 61070 -- 61098), the accretion dynamics shifted dramatically. The emergence of a strong multi-color disk blackbody component ($T_{\rm in} = 0.88 \pm 0.06$ to $1.23 \pm 0.10$ keV) signals that the optically thick, geometrically thin accretion disk had migrated inward, likely reaching the innermost stable circular orbit (ISCO). Concurrently, the corona cooled and the photon index softened significantly to $\Gamma = 2.45 \pm 0.25$ to $2.51 \pm 0.15$. This behavior is indicative of strong cooling of the coronal plasma due to the high flux of soft thermal photons emitted by the inner disk.

The transition from the hard state to the soft state is marked by the hardening of the disk blackbody component and the softening of the photon index ($\Gamma$). Within the Two-Component Advective Flow (TCAF) framework \citep{Chakrabarti1995}, this spectral evolution is driven by the dynamic shift in the relative mass accretion rates between the Keplerian disk and the sub-Keplerian halo. The findings of this report suggest that as the Keplerian disk rate increases during the outburst rise, it forces the shock location inward, flooding the inner region with soft seed photons and naturally cooling the hot Comptonizing plasma.

\subsection{Black Hole Spin and System Geometry}
\label{subsec:spin_geometry}
The bright soft state provided high signal-to-noise spectra, allowing for precise constraints on the relativistic reflection features, namely the broadened Fe K$\alpha$ line and the Compton hump. With the unmodeled instrumental soft-excess formally resolved, the reflection modeling yields a high dimensionless spin parameter ranging between $a \approx 0.94 - 0.99$. This finding is in excellent agreement with previous analyses of the 2015 outburst of GS 1354-64 \citep[e.g.,][]{ElBatal2016}, as well as the independent broadband modeling of the early 2026 hard state using high-resolution \textit{XRISM} data by \citet{Liu2026}, who also reported a rapidly spinning black hole ($a_* > 0.98$). The consistency of this measurement across different outbursts separated by a decade underscores the reliability of X-ray reflection spectroscopy in measuring fundamental black hole parameters when the disk extends to the ISCO.

Furthermore, through our joint spectral modeling, the inclination of the inner accretion disk was tightly constrained and remarkably stable across the active epochs, evaluating to $i = 51.55^\circ \pm 0.17^\circ$. This moderate inclination is consistent with the observable broadening of the iron line profile due to strong Doppler and relativistic effects, and aligns well with prior optical and X-ray estimations of the binary system's orbital plane.

\subsection{Highly Ionized Disk Winds in the Hard State}
\label{subsec:disk_winds}
One of the most striking results of this analysis is the detection of deep absorption features at $\sim 7.3$ keV during the rising hard and hard-intermediate states (MJD 61053, 61061, and 61064). Modeled successfully with a Gaussian absorption component (\texttt{gabs}), this feature is robustly interpreted as the presence of a highly ionized, outward-flowing disk wind, likely dominated by hydrogen-like Iron (Fe XXVI).

In standard black hole X-ray binaries, equatorial disk winds are predominantly observed during the disk-dominated soft states and are typically quenched during the hard states, where compact steady jets dominate the mass outflow \citep{Ponti2012}. The detection of a highly ionized wind during the hard state of GS 1354-64 is therefore highly notable. Given the system's inclination (constrained between $45^\circ$ and $53^\circ$ across the individual epochs), our line of sight likely intercepts the upper boundary of an equatorial outflow. The presence of this wind in the hard state may suggest the presence of a strong localized radiation pressure before the global transition to the soft state occurred. Similar anomalous hard-state winds have been observed in a handful of high-inclination systems (e.g., V404 Cygni; \citealt{King2015}), highlighting the complex interplay between jet launching and wind driving mechanisms.

However, it is crucial to note that an independent analysis of the concurrent hard state (MJD 61060--61063) utilizing the high-resolution microcalorimeter aboard \textit{XRISM} by \citet{Liu2026} reported a clean, broad iron line without any significant narrow absorption features. The discrepancy between this report's \textit{NuSTAR} detection and the \textit{XRISM} non-detection presents an intriguing puzzle. This could imply that the disk wind is highly variable and transient on short timescales. Alternatively, the lower spectral resolution of NuSTAR may be blending complex reflection substructures, artificially mimicking a discrete absorption line that the superior resolution of XRISM resolves into a clean continuum. To definitively break this degeneracy and map the true ionization structure of the outflow, a comprehensive joint modeling of these simultaneous NuSTAR and XRISM datasets is beyond the scope of this work, and will instead be presented in a dedicated follow-up paper.

\subsection{Outburst Decay and the Fading of Reflection}
\label{subsec:decay_degeneracy}
As the outburst entered its decay phase (MJD 61117), the overall X-ray flux dropped substantially, and the source began migrating back toward the intermediate/hard state, evidenced by the hardening of the continuum ($\Gamma = 2.04 \pm 0.27$) and the cooling of the thermal disk ($T_{\rm in} = 0.86 \pm 0.06$ keV).

During this faint phase, the relativistic reflection features that prominently anchored the spectral modeling during the outburst peak faded entirely into the background noise. Rather than forcing a statistically degenerate relativistic reflection model (\texttt{relxill}) onto low-luminosity data---which artificially warps structural parameters to minimize fit statistics against Poisson noise---a simpler \texttt{diskbb + powerlaw} model was adopted. This physically captures the system's transition: as the mass accretion rate plummeted, the dense coronal structure responsible for the strong Compton back-scattering dissipated, leaving behind a cooling thermal disk and a weakening, hardening primary power-law. This macroscopic structural shift unambiguously confirms the quenching of the accretion flow as GS 1354-64 returned to quiescence.

\subsection{Coronal Geometry: Joint Spectral and Polarimetric Constraints}
\label{subsec:coronal_geometry}
The IXPE observation (MJD 61078--61083) provides a critical, model-independent test of the spectral findings. Remarkably, this IXPE exposure strictly overlaps with the \textit{NuSTAR} observation 91202311002 (MJD 61082.85). During this exact epoch, the \texttt{relxill} spectral modeling indicated the source was in a strongly disk-dominated soft state ($T_{\rm in} = 0.89 \pm 0.06$ keV) with a cooled, soft corona ($\Gamma = 2.49 \pm 0.23$) and an inner disk inclination of $i = 48.0^\circ \pm 3.3^\circ$.

The IXPE polarimetry from this simultaneous window corroborates and geometrically extends this picture. The joint spectro-polarimetric modeling explicitly quantifies this structure via the linear tracking parameters. The polarization slope parameter reveals a distinct, highly significant positive trend ($A_{\rm slope} = 1.3\% \pm 0.2\%$ per keV), challenging an isotropic or flat energy-independent coronal profile. The rising polarization degree with energy—culminating at $\sim 7.5\% - 9.1\%$ in the 6--8 keV band—suggests that the Comptonizing corona is unlikely to be a spherically symmetric cloud, which would yield zero net polarization. Instead, to achieve such a high degree of linear polarization at an intermediate viewing angle (constrained between $\sim 45^\circ$ and $55^\circ$), the coronal plasma is likely highly asymmetric and radially extended. These joint spectral and polarimetric measurements provide compelling evidence for a flat, ``slab'' or ``wedge'' like coronal geometry that radially sandwiches the inner accretion disk. High-energy photons likely undergo multiple scattering events within the plane of this extended slab before escaping towards the observer, thereby acquiring the high degree of polarization observed in the harder bands.

Furthermore, the strict energy-independence of the polarization angle (which remains remarkably stable near $-2^\circ$ across all bands) isolates the geometric axis of the corona. Because the thermal accretion disk is intrinsically unpolarized in this regime, it does not contribute to the net polarization angle. Thus, this constant PA suggests that the highly polarized, extended slab corona maintains a single, stable axis of symmetry, strongly disfavoring any significant macroscopic precessional wobbling or misalignments in the inner scattering flow during the soft state.

To evaluate the true coronal geometry, I consider three primary scattering scenarios. A spherically symmetric corona is disfavored, as the cancellation of orthogonal vectors would yield a negligible net polarization. We then test the lamppost model; while popular, this geometry generally struggles to reproduce both the high $PD$ ($\sim 8 - 9\%$) and the strict energy-independence of the $PA$ simultaneously, often requiring fine-tuned viewing angles or extreme scattering conditions that are not supported by the reflection fits. In contrast, the radially extended slab/sandwich geometry provides a natural explanation: the disk-parallel symmetry naturally aligns the polarization vectors, resulting in the constant $PA$ near $-2^\circ$. Furthermore, the observed energy-dependent increase in $PD$ is a characteristic signature of multiple electron scattering events occurring within an extended, plane-parallel medium.

While the polarimetric results of this report aim to constrain the corona to a radially extended, asymmetric slab geometry, this morphology is physically consistent with theoretical models such as TCAF \citep{Chakrabarti1995, Mondal2024}. The CENtrifugal pressure supported BOundary Layer (CENBOL), which TCAF identifies as the primary Compton cloud, can naturally manifest as such a highly flattened, slab-like geometry when the centrifugal support is compressed by strong Compton cooling from the underlying disk. The IXPE data thus provides an independent, energy-resolved geometric map of the CENBOL’s structure, supporting the idea that in the soft state, it is not a symmetric cloud, but rather an extended medium that tightly sandwiches the inner accretion disk.

For the sake of completeness, I note that while the empirical spectro-polarimetric modeling favors this extended slab geometry, executing full general relativistic Monte Carlo radiative transfer simulations (utilizing codes such as MONK) would provide further theoretical validation. Such simulations could more precisely map the observed energy-dependent polarization signatures to specific electron temperature and optical depth profiles within the CENBOL. While beyond the scope of the current observational study, applying these advanced radiative transfer codes offers a compelling avenue for future theoretical investigations of this source.

\subsection{The Physical Origin of the LFQPOs and Disk-Corona Coupling}
\label{sec:qpo_coupling}

\begin{figure}[!ht]
    \centering
    \includegraphics[width=0.42\textwidth,height=6cm]{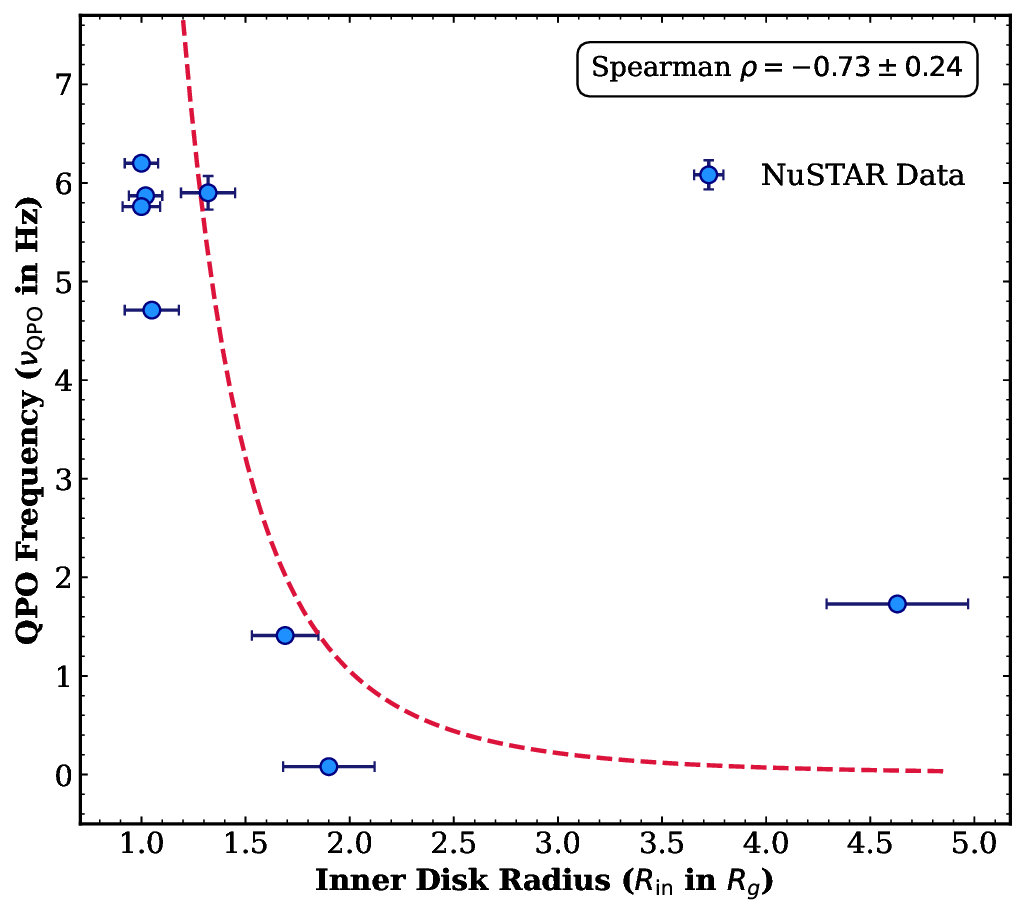}
    \includegraphics[width=0.44\textwidth,height=6cm]{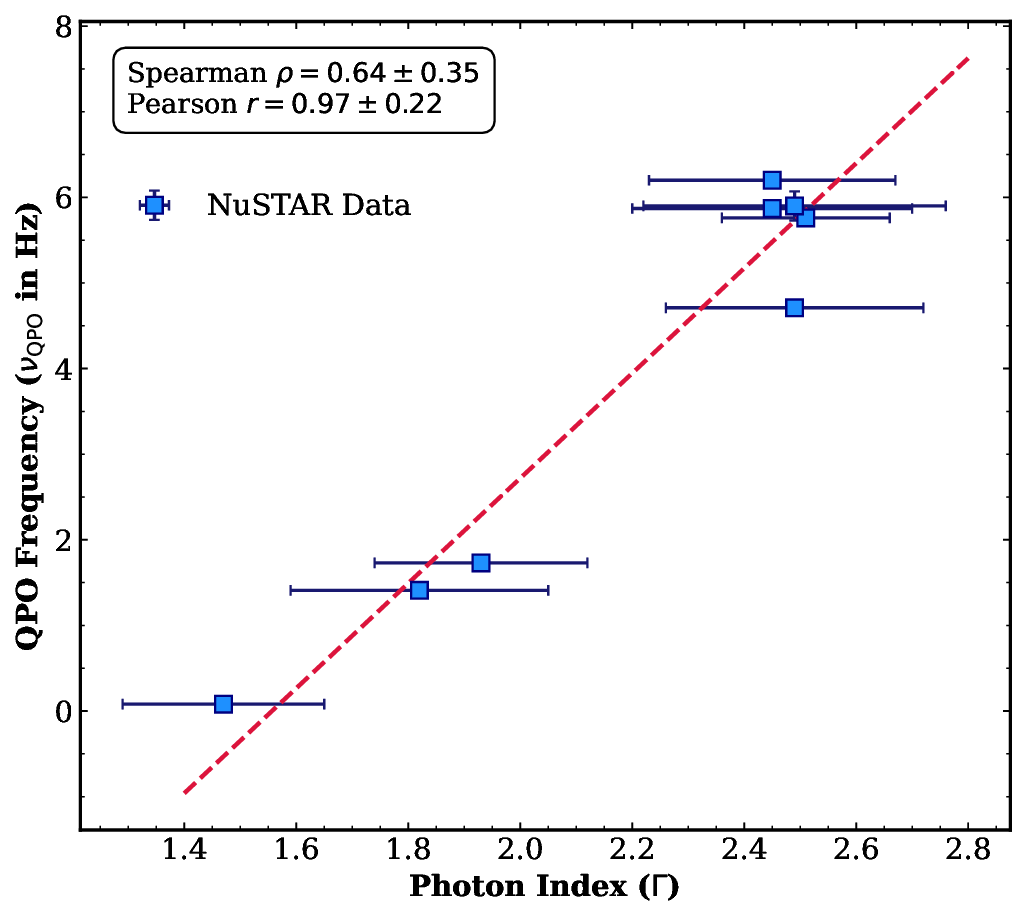}
    \caption{\textit{Top:} The evolution of the QPO centroid frequency ($\nu_{\mathrm{QPO}}$) as a function of the reflection-derived inner disk radius ($R_{\mathrm{in}}$). The strong anti-correlation (Spearman $\rho \approx -0.73 \pm 0.22$) suggests a migrating inner flow. \textit{Bottom:} The QPO frequency correlated with the coronal photon index ($\Gamma$). The linear coupling demonstrates that as the disk migrates inward (driving the QPO frequency up), the increased flux of soft seed photons cools the corona (driving $\Gamma$ up).}
    \label{fig:qpo_correlations}
\end{figure}

To determine the physical mechanism driving the low-frequency QPOs during the hard-to-soft transition of GS 1354-64, I investigated the coupling between the timing properties and the evolving spectral geometry. In the context of the Relativistic Precession Model (RPM; \citealt{Stella1998}), Type-C LFQPOs are generated by the Lense-Thirring precession of the hot inner flow (the corona) misaligned with the black hole spin axis. Alternatively, within the Two-Component Advective Flow (TCAF) framework \citep{Chakrabarti1995}, the LFQPOs can be interpreted as the physical oscillation of the shock boundary itself. In this scenario, the shock that defines the outer edge of the CENBOL becomes unstable when the cooling timescale---modulated by the influx of soft seed photons from the advancing Keplerian disk---matches the dynamical timescale of the shock.

As shown in the left panel of Figure \ref{fig:qpo_correlations}, I find a robust monotonic anti-correlation (Spearman rank $\rho = -0.73 \pm 0.24$) between the measured QPO centroid frequencies and the inner disk radii derived independently from the \texttt{relxill} reflection modeling. This scaling relation closely aligns with the expectations of both the RPM and TCAF frameworks, supporting the scenario that the inward migration of the accretion disk edge toward the ISCO is the primary geometric driver increasing the characteristic frequency of the flow. Whether through the increase of the Lense-Thirring precessional rate or the dynamic compression of the CENBOL shock front as it is squeezed by the advancing Keplerian disk, the inward drift of the disk appears to strongly drive the frequency evolution.

Furthermore, this geometric evolution leaves a distinct signature on the thermal state of the corona. The right panel of Figure 14 demonstrates a strong linear correlation between $\nu_{\mathrm{QPO}}$ and the coronal photon index ($\Gamma$), supported by a Pearson coefficient of $r = 0.97 \pm 0.22$ and a Spearman rank of $\rho = 0.64 \pm 0.35$. As the accretion disk moves inward (yielding higher QPO frequencies), it penetrates deeper into the hot inner flow. This structural change likely floods the corona with a higher flux of soft thermal seed photons from the disk, leading to highly efficient inverse-Compton cooling. This cooling mechanism manifests as the observed steepening (softening) of the photon index.

\begin{figure}[ht!]
    \centering
    \begin{subfigure}[b]{0.48\textwidth}
        \centering
        \includegraphics[width=0.9\textwidth,height=6cm]{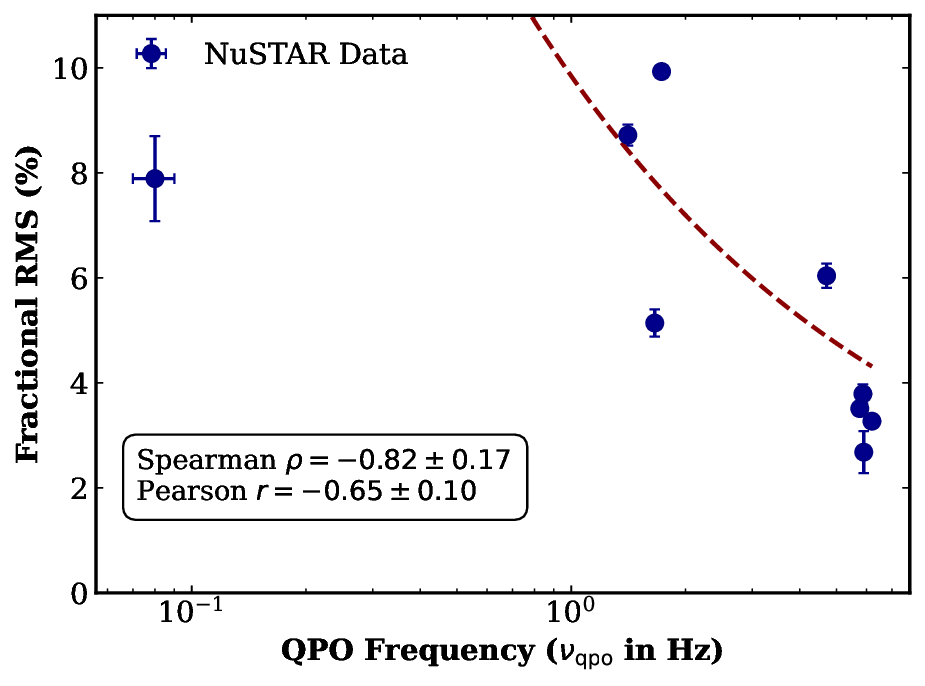}
    \end{subfigure}\hfill
    \begin{subfigure}[b]{0.48\textwidth}
        \centering
        \includegraphics[width=0.9\textwidth,height=6cm]{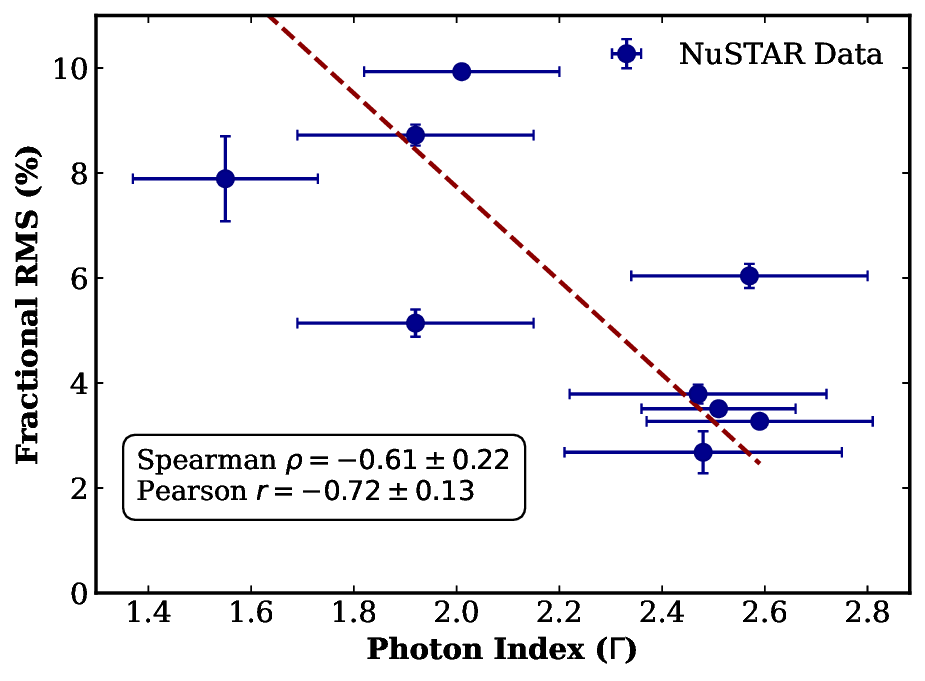}
    \end{subfigure}
	\caption{Correlations between the QPO fractional RMS and the evolution of the accretion geometry. \textit{Top:} The strong anti-correlation ($r \approx -0.65 \pm 0.10$; $\rho \approx -0.82 \pm 0.17$) between QPO frequency and RMS amplitude tracks the physical collapse of the precessing or oscillating inner flow. \textit{Bottom:} The significant anti-correlation ($r \approx -0.72 \pm 0.13$, $\rho \approx -0.61 \pm 0.22$) between the coronal photon index ($\Gamma$) and fractional RMS suggests that as the accretion disk migrates inward and cools the corona, the emitting volume participating in the Lense-Thirring precession or shock oscillation is effectively reduced.}
    \label{fig:rms_correlations}
\end{figure}

To further support these geometric and hydrodynamic models, I analyze the evolution of the fractional RMS of the fundamental QPO. As shown in Figure \ref{fig:rms_correlations}, the QPO amplitude exhibits strong negative correlations with both the centroid frequency and the photon index. The observed trends indicate that as the inner flow migrates to smaller radii---thereby increasing the characteristic precessional or shock oscillation frequency---the fractional RMS amplitude drops significantly. Similarly, the anti-correlation between $\Gamma$ and RMS provides independent evidence for the disk-cooling mechanism: as the disk advances inward, the increasing seed photon flux cools the corona, leading to spectral softening and a concomitant reduction in the precessing or oscillating coronal volume. Collectively, these results support a scenario where the geometric collapse of the CENBOL and the quenching of the large-scale oscillations are driven by the inward encroachment of the accretion disk.

\subsection{Physical Implications: A Unified Spectro-Temporal-Polarimetric View}

\begin{figure}[ht!]
    \centering
    \includegraphics[width=0.45\textwidth,height=8.3cm]{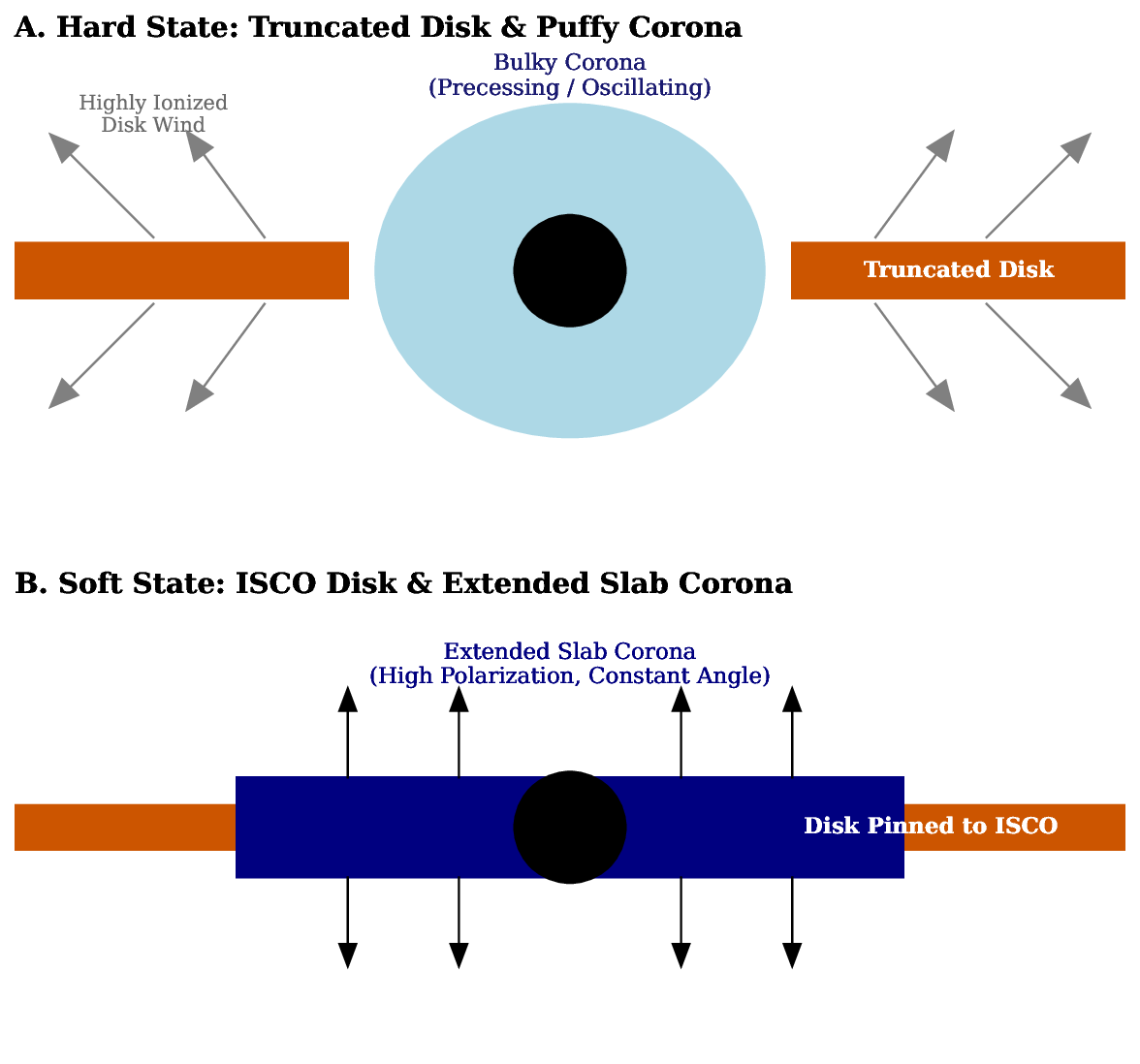}
    \caption{A physical schematic illustrating the geometrical evolution of the inner accretion flow in GS 1354-64 across its outburst states. \textbf{(A) Hard State:} The standard accretion disk is truncated at a larger radius. The inner region is dominated by a bulky, quasi-spherical corona (representing an oscillating CENBOL or precessing flow). This massive structure drives the high-RMS Type-C QPOs and launches the highly ionized equatorial disk winds observed by \textit{NuSTAR}. \textbf{(B) Soft State:} The accretion disk migrates inward and pins to the ISCO ($a > 0.99$). The corona collapses into a highly asymmetric, radially extended slab that sandwiches the inner disk. Scattering within this dense horizontal plane produces the steep $8-9\%$ polarization degree observed by \textit{IXPE}. The suppression of macroscopic wobbling in this stable configuration naturally enforces the strictly constant polarization angle and quenches the high-RMS QPOs.}
    \label{fig:geometry_schematic}
\end{figure}

The findings of this report synthesize the spectral, timing, and polarimetric properties into a unified framework. While the TCAF model provides the dynamic mechanism---where the advancing Keplerian disk dictates the shock geometry and cools the corona---the polarimetric data provides the geometric map of this process. The suppression of the high-RMS macroscopic Type-C QPOs during the soft state signals the breakdown of large-scale Lense-Thirring precession. The flow stabilizes, and the CENBOL flattens into the extended, symmetric slab strongly supported by this report's \textit{IXPE} constraints. This suggests that the hard-to-soft transition in GS 1354-64 is not merely a change in spectral hardness, but a substantial structural reconfiguration of the accretion flow driven by the interaction between the migrating disk and the centrifugal barrier.

A transition from a bulky, dynamically active flow to a stable, radially extended structure elegantly resolves the observed geometric and timing behaviors. As the standard Keplerian disk migrates to the ISCO, the large-scale Lense-Thirring precession that dominated the rising hard state is quenched, causing the Comptonizing plasma to settle into a symmetric slab geometry sandwiching the inner disk. The high density of in-plane scatterings within this flattened wedge naturally produces the steep, energy-dependent polarization degree peaking at $\sim 7.5\%-9.1\%$, while its stable axis of symmetry maintains the consistent polarization angle. The residual weak, high-frequency timing features observed during this soft epoch can then be attributed to localized radial boundary oscillations or breathing modes rather than global, macroscopic rotational wobbling. A conceptual representation of this global geometric and physical restructuring across the outburst states is illustrated in Figure \ref{fig:geometry_schematic}.

\section*{Summary and Conclusions}
\label{sec:summary}

In this paper, I presented a comprehensive multi-wavelength analysis of the highly active 2025--2026 outburst of the dynamically confirmed black hole candidate GS 1354-64. 
By leveraging simultaneous broadband spectroscopy (\textit{NuSTAR}), fast timing, and X-ray polarimetry (\textit{IXPE}), I tracked the evolution of the accretion flow across its state transitions. 
The primary conclusions drawn from this investigation are summarized as follows:

\begin{enumerate}

    \item \textbf{Outburst Evolution and Relativistic Reflection:} The source successfully transitioned from a Compton-dominated hard state to a luminous, thermal-dominated soft state. Initial reflection modeling of the individual epochs constrained the black hole to be rapidly spinning ($a_* \approx 0.94 - 0.99$) with an inner disk inclination of $i \approx 45^\circ - 53^\circ$. Subsequent joint spectral modeling across these states successfully eliminated residual degeneracies, tightly anchoring the near-maximal spin ($a_* \approx 0.998$) and strictly constraining the inclination to $i = 51.55^\circ$, confirming the disk is rigidly pinned to the ISCO.

    \item \textbf{Anomalous Hard-State Disk Winds:} During the rising hard and hard-intermediate states, I detected significant absorption features at $\sim 7.3$ keV, indicating the presence of a highly ionized (Fe XXVI), outward-flowing equatorial disk wind prior to the soft-state transition. 
    This suggests the accretion rate transiently reached locally near-Eddington limits in the inner regions during the outburst rise. 
    
    \item \textbf{An Extended, Asymmetric Corona:} Joint spectro-polarimetric modeling of the simultaneous soft-state exposure revealed a steep, energy-increasing polarization degree, peaking at $\sim 7.5\% - 9.1\%$ in the 6--8 keV band. 
    This strongly disfavors spherically symmetric or compact lamppost models, providing compelling evidence for a radially extended, highly asymmetric ``slab" or ``wedge" geometry. 
    This morphology is physically consistent with the Two-Component Advective Flow (TCAF) framework, where the CENBOL flattens in response to intense cooling from the underlying disk. 

    \item \textbf{Evolution and Physical Origin of the LFQPOs:} Fast timing analysis revealed a clear dynamical evolution from strong, highly coherent Type-C QPOs (0.08--1.73 Hz, RMS $\sim 8-10\%$) in the rising hard state to weak, broad features (4.7--6.2 Hz, RMS $\sim 3-6\%$) in the soft state. 
Crucially, I discovered a strong anti-correlation between the QPO frequency and the inner disk radius ($R_{\mathrm{in}}$), as well as a linear correlation with the photon index ($\Gamma$). 
These scaling relations support the scenario that the inward migration of the truncated disk drives the macroscopic precession of the hot inner flow while simultaneously cooling it. 

    \item \textbf{A Unified Stable Geometry:} The physical tension between a macroscopically precessing corona and a consistently stable polarization angle is elegantly resolved by this unified dataset. 
The quenching of the high-RMS macroscopic QPO precisely when the polarization angle is observed to be locked near $-2^\circ$ suggests that the corona stabilized into an extended slab. 
The weak, high-frequency residual QPOs observed during this phase are likely driven by localized geometric perturbations (such as radial boundary oscillations) rather than large-scale rotational wobbling. 

\end{enumerate}

Ultimately, the 2025--2026 outburst of GS 1354-64 demonstrates the diagnostic power of combining high-resolution X-ray timing, broadband spectroscopy, and polarimetry to map coronal geometry. 
These joint constraints suggest that the soft-state corona in this system is a highly stable, radially extended structure.

\section*{Data Availability}
\label{sec:data_avail}

The raw, uncalibrated X-ray data from the \textit{NuSTAR} and \textit{IXPE} observatories utilized in this manuscript are completely public and can be freely accessed through the High Energy Astrophysics Science Archive Research Center (HEASARC) data archive, maintained by NASA/GSFC (\url{https://heasarc.gsfc.nasa.gov/}). The specific observation IDs, alongside their exact exposure times covering the 2025--2026 outburst of GS 1354-64, are explicitly listed in Table \ref{tab:obs_log}. 

The data reduction pipelines, calibration databases (CALDB), and software suites used in this work—namely the HEASoft \texttt{FTOOLS} and the \texttt{IXPEOBSSIM} package—are open-source and publicly available from their respective official repositories. Any additional derived higher-level data products, folded spectra, or custom analysis scripts supporting the findings of this study are available from the corresponding author upon reasonable request.

\section*{Acknowledgements}
\label{sec:ack}

I acknowledge support from the 2026 National Natural Science Foundation of China (NSFC) RFIS-I Grant (W2633019), the 79th General Grant the China Postdoctoral Science Foundation (Certificate Number: 2026M793513), and CaiYun-2024 Postdoctoral Grant for Innovation Project of Yunnan Province (C615300504128). I also thank the referee(s) for the fruitful comments, which made this manuscript scientifically better.

\section*{Facilities and Softwares}
\label{sec:facilities}

\textbf{Facilities:} \textit{NuSTAR} (NASA/Caltech/JPL), \textit{IXPE} (NASA/ASI).

\textbf{Software:} This research made use of the \texttt{HEASoft} suite (\textbf{v6.35.1}) of data reduction and analysis tools, including \texttt{FTOOLS} and \texttt{XSPEC} (\textbf{v12.15.0}) \citep{Arnaud1996}, provided by the High Energy Astrophysics Science Archive Research Center (HEASARC). X-ray polarimetric data processing and model-independent analysis were performed using the \texttt{IXPEOBSSIM} software suite (v31.0.3) \citep{Baldini2022}. Spectral modeling heavily utilized the \texttt{relxill} relativistic reflection package (v2.3) \citep{Garcia2014, Dauser2014}. Custom timing analysis, Power Density Spectrum (PDS) multi-Lorentzian fitting, and graphical visualizations were executed in a \texttt{Python} (v3.10) environment utilizing the \texttt{NumPy}, \texttt{SciPy}, and \texttt{Matplotlib} libraries.

\bibliographystyle{elsarticle-harv}

\onecolumn
\appendix

\section{Resolving the Dual-Peak QPO Structure}
\label{sec:appendix_qpo_split}

During the rapidly evolving hard-intermediate state observation on MJD 61061.54 (ObsID 91201302004), the source exhibited a complex fundamental QPO structure. In this appendix, I demonstrate the statistical necessity of employing a dual-Lorentzian model to accurately capture this feature without leaving structured residuals. To statistically validate the presence of the dual-peak structure, an F-test was performed comparing the single-Lorentzian and dual-Lorentzian models. The inclusion of the second Lorentzian component significantly improved the fit from $\chi^2/d.o.f. = 576.87/247$ to $\chi^2/d.o.f. = 356.75/244$. This yielded an F-statistic of 50.18 with a null hypothesis probability of $2.68 \times 10^{-25}$, unequivocally confirming the physical necessity of the dual-peak QPO model during this highly dynamic transition phase.

\begin{figure}[ht!]
    \centering
        \centering
    \begin{subfigure}[b]{0.5\textwidth}
        \centering
        \includegraphics[width=\textwidth]{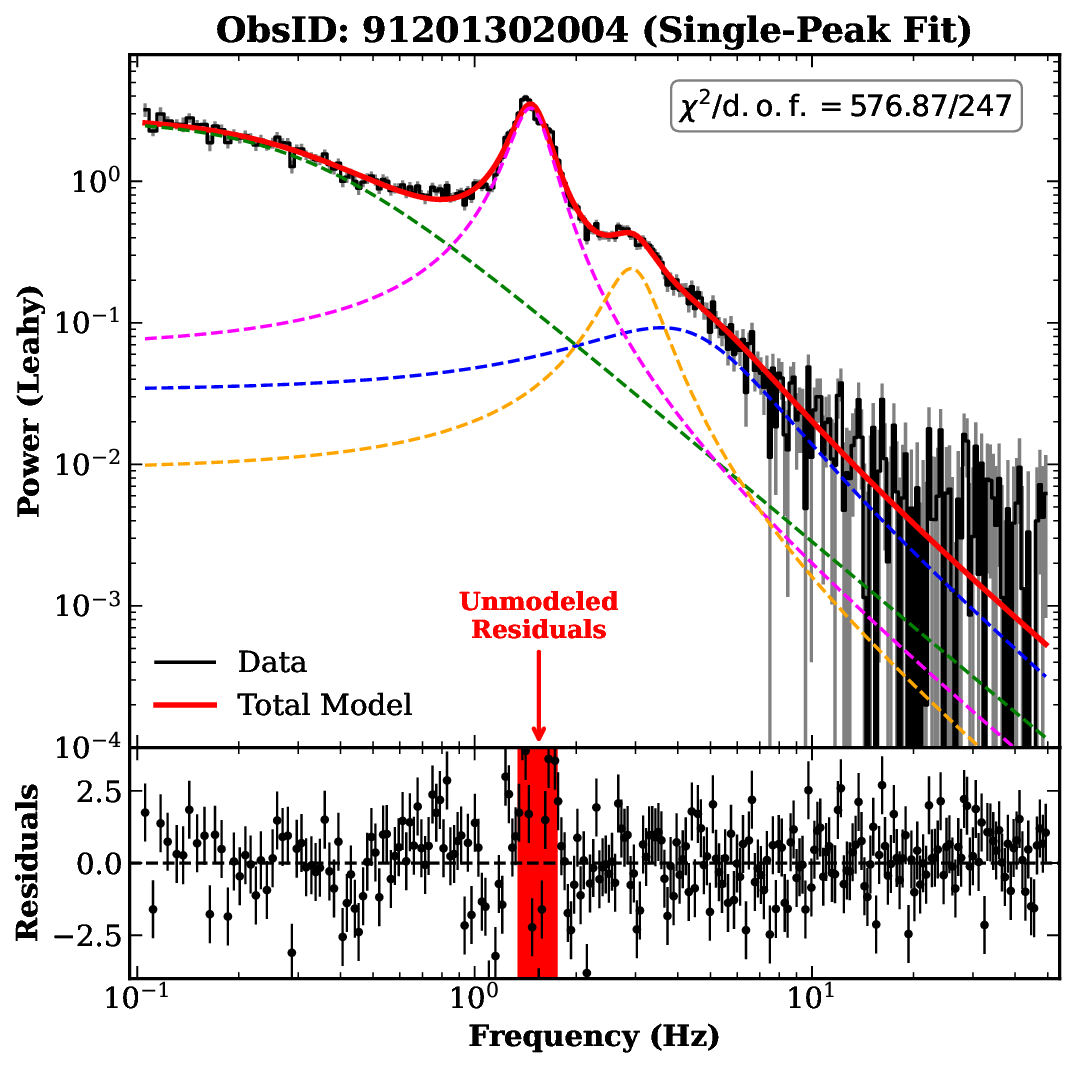}
    \end{subfigure}\hfill
    \begin{subfigure}[b]{0.5\textwidth}
        \centering
        \includegraphics[width=\textwidth]{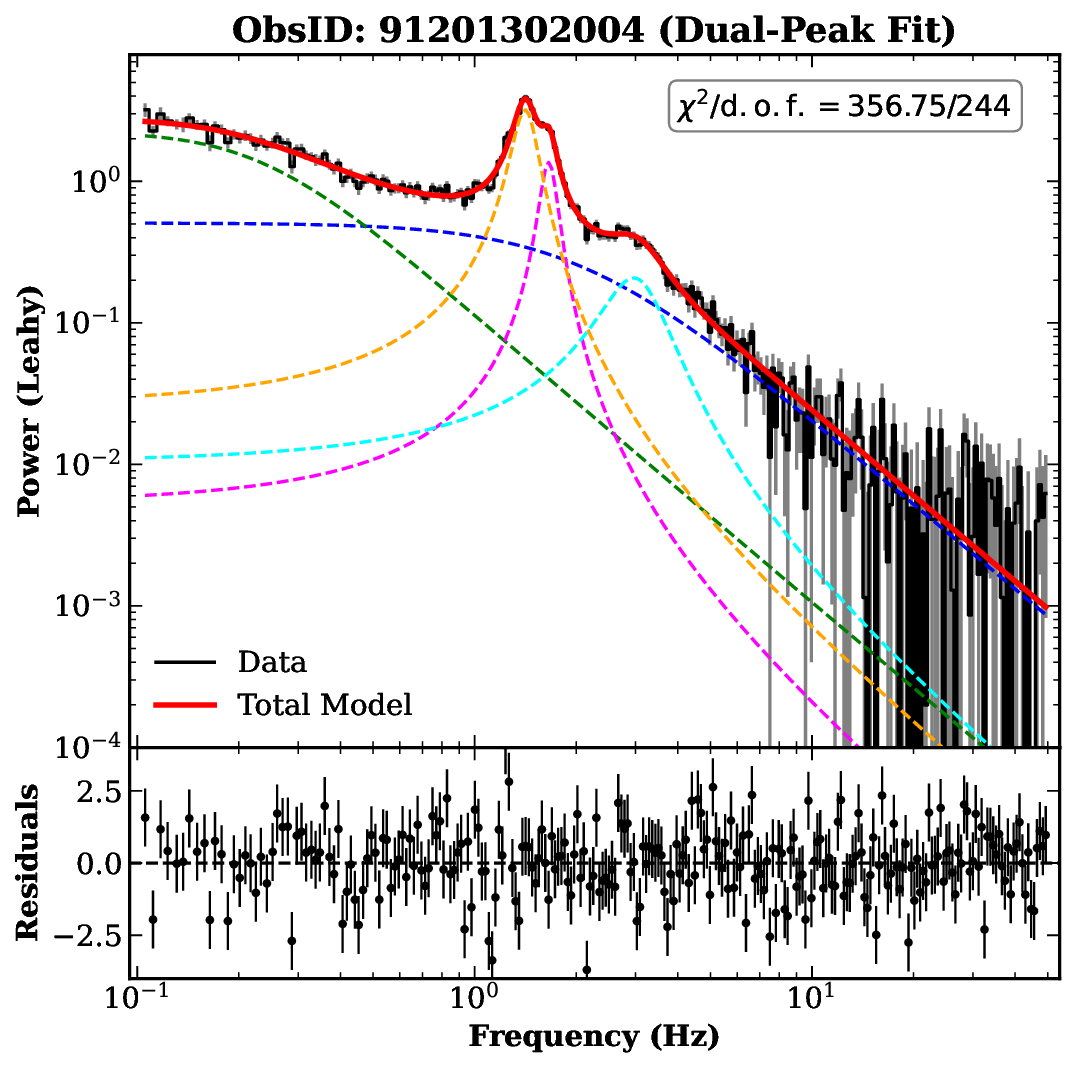}
    \end{subfigure}\hfill
    \caption{Comparison of time-averaged cospectrum modeling for ObsID 91201302004. \textbf{Left:} A single-Lorentzian fit to the fundamental QPO leaves significant, structured wave-like residuals between $1.4$ and $1.7$ Hz, indicating an inadequate model for the rapidly evolving precessional flow. \textbf{Right:} The inclusion of a second, adjacent narrow Lorentzian profile ($\sim 1.41$ Hz and $\sim 1.66$ Hz) successfully flattens the residuals ($\Delta\chi$), achieving a statistically robust fit. This confirms the dual-peak structure of the QPO during this highly dynamic transition phase.}
    \label{fig:qpo_split_comparison}
\end{figure}

\section{Dynamical Cospectra of the Soft State}
\label{sec:appendix_soft_state}

In this appendix, I present the dynamical cospectra for the soft state observations to visually demonstrate the suppression of the macroscopic Quasi-Periodic Oscillations following the state transition.

\begin{figure*}[ht!]
    \centering
    \begin{subfigure}[b]{0.33\textwidth}
        \centering
        \includegraphics[width=\textwidth]{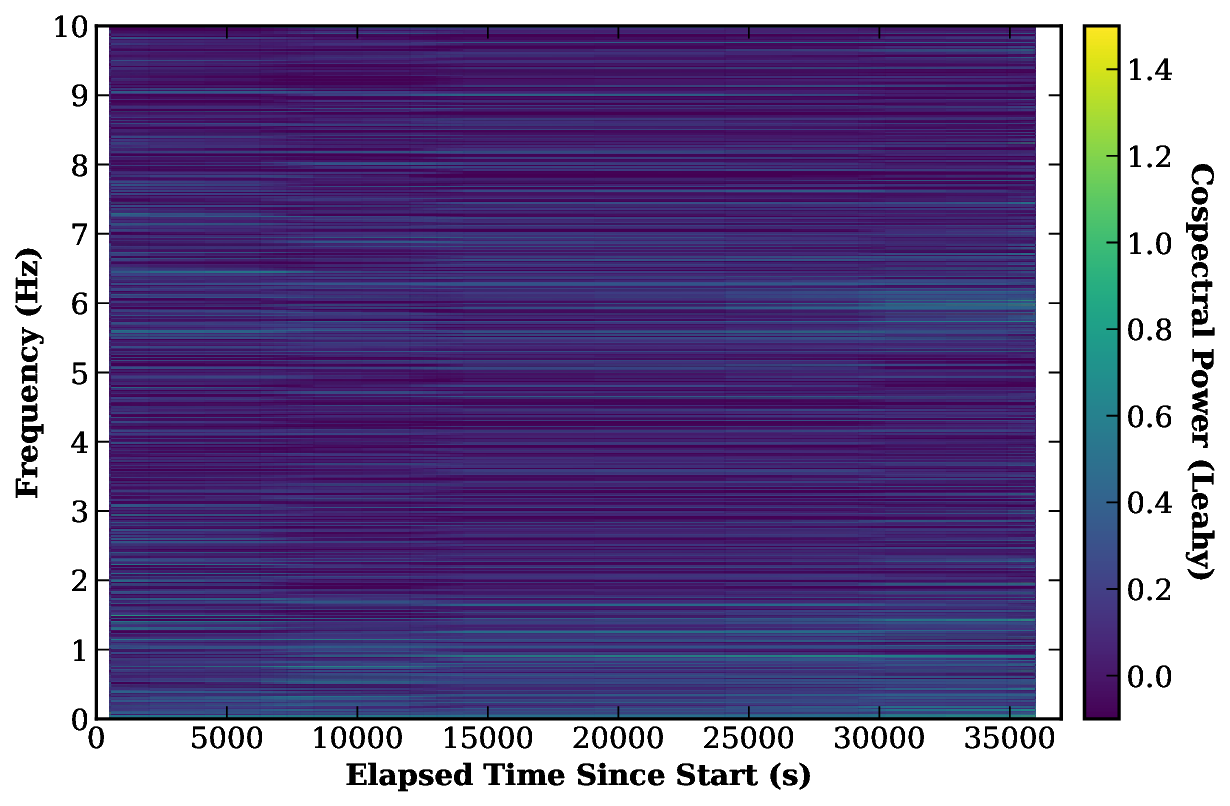}
    \end{subfigure}\hfill
    \begin{subfigure}[b]{0.33\textwidth}
        \centering
        \includegraphics[width=\textwidth]{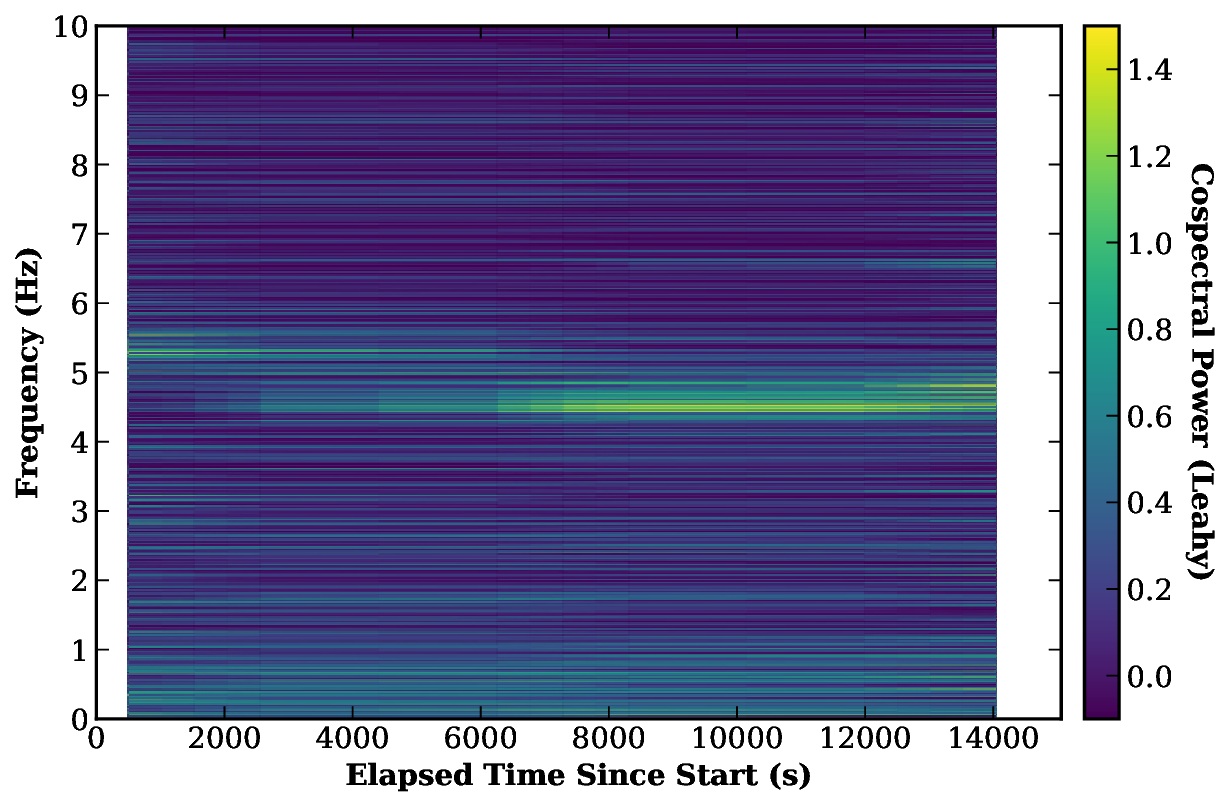}
    \end{subfigure}\hfill
    \begin{subfigure}[b]{0.33\textwidth}
        \centering
        \includegraphics[width=\textwidth]{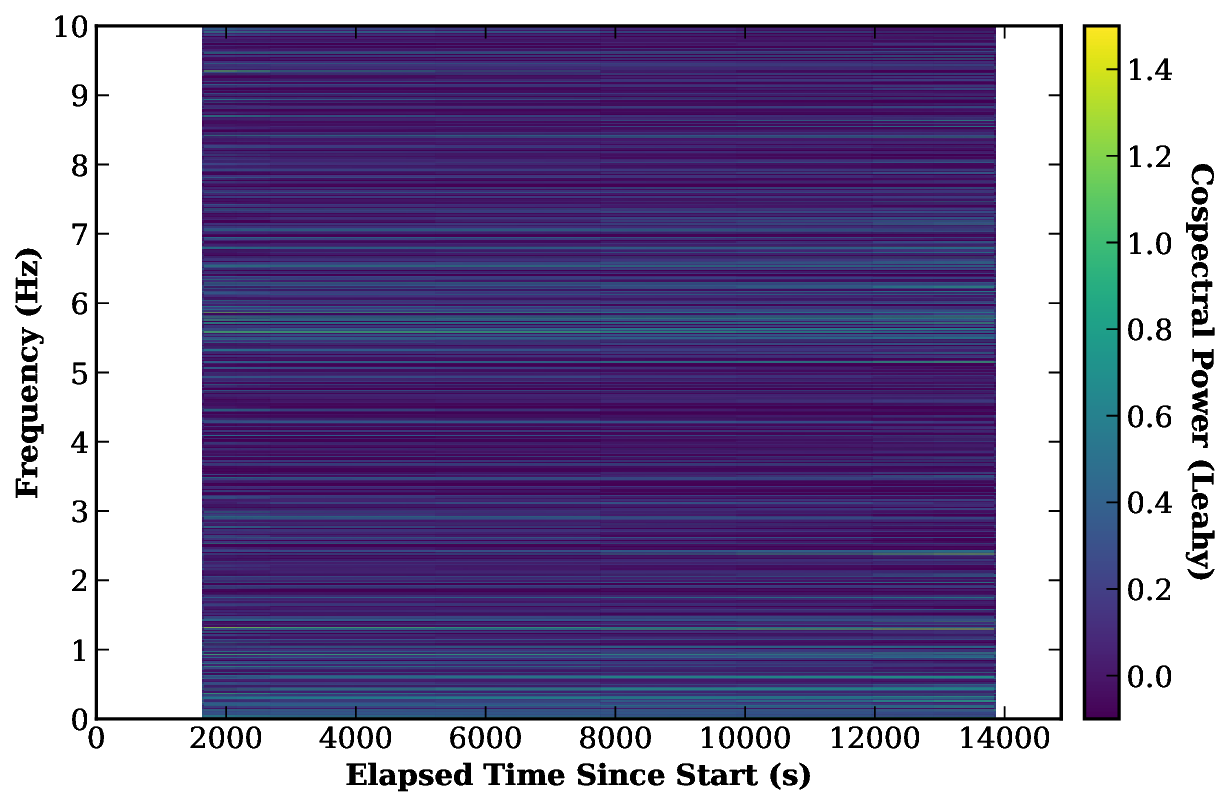}
    \end{subfigure}
    \caption{\textit{NuSTAR} dynamic cospectra during the luminous soft state of GS 1354--64. \textbf{Left:} MJD 61070.33 (ObsID 91202308002). \textbf{Center:} MJD 61082.85 (ObsID 91202311002). \textbf{Right:} MJD 61085.17 (ObsID 91202311004). Unlike the bright precessional tracks observed during the hard state, these plots appear featureless. Because Fourier power scales with the square of the fractional RMS, the collapse of the inner corona causes the QPO signal to drop by nearly an order of magnitude. Consequently, the peak amplitude of the signal in any single 1024-second snapshot falls completely below the random statistical variance (the Poisson noise floor) inherent to short exposures, physically necessitating the full time-averaged cospectra (Table \ref{tab:qpo_parameters}) to extract the faint, high-frequency boundary oscillations.}
    \label{fig:soft_state_dynamic}
\end{figure*}

\end{document}